\documentstyle[12pt,axodraw,epsfig,url]{article}

\newcommand\hepnumber{hep-ph/0512128}

\def\Title#1{\begin{center} {\Large #1 } \end{center}}
\def\Author#1{\begin{center}{ \sc #1} \end{center}}
\def\Address#1{\begin{center}{ \it #1} \end{center}}

\newcommand{\arline}{\nonumber \\}
\newcommand\pubblock{\rightline{\begin{tabular}{r} 
         \today \\ \hepnumber \end{tabular}}}
\newenvironment{Abstract}{\begin{quotation} \begin{center}
                       ABSTRACT
     \end{center}\bigskip  }{\end{quotation}}

\def\Tr{\,{\rm Tr}\,}

\def\eps{\epsilon}
\def\tSigma{\tilde{\Sigma}}
\def\half{\frac{1}{2}}
\def\mT{m_{{\rm T}_+}}
\def\psip{\psi^\prime}

\def\beq{\begin{equation}}
\def\eeq#1{\label{#1}\end{equation}}
\def\eeqn{\end{equation}}
\def\beqa{\begin{eqnarray}}
\def\eeqa#1{\label{#1}\end{eqnarray}}
\def\eeqan{\end{eqnarray}}
\def\CR{\nonumber \\ }
\def\leqn#1{(\ref{#1})}
\newcommand\iden{\leavevmode\hbox{\small1\normalsize\kern-.33em1}}

\def\Cornell{Institute for High-Energy Phenomenology, \\ 
Newman Laboratory for Elementary Particle Physics, \\
Cornell University, Ithaca, NY~14853 USA} 

\def\stacksymbols #1#2#3#4{\def\theguybelow{#2}
    \def\vp{\lower#3pt}
    \def\sp{\baselineskip0pt\lineskip#4pt}
    \mathrel{\mathpalette\intermediary#1}}

\def\intermediary#1#2{\vp\vbox{\sp
     \everycr={}\tabskip0pt
     \halign{$\mathsurround0pt#1\hfil##\hfil$\crcr#2\crcr
              \theguybelow\crcr}}}

\def\gapproxeq{\stacksymbols{>}{\sim}{2.5}{.2}}
\def\lapproxeq{\stacksymbols{<}{\sim}{2.5}{.2}}

\begin{document}
\begin{titlepage}
\pubblock

\vfill
\Title{Little Higgs Models and Their Phenomenology\footnote{To be published 
in {\it Progress of Particle and Nuclear Physics 2006.}}}
\vfill
\Author{Maxim Perelstein}
\Address{\Cornell}
\vfill
\begin{Abstract}
This article reviews the Little Higgs models of electroweak symmetry 
breaking and their phenomenology. Little Higgs models incorporate a light 
composite Higgs boson and remain perturbative until a scale of order 10 TeV, 
as required by precision electroweak data. The collective symmetry breaking 
mechanism, which forms the basis of Little Higgs models, is introduced. 
An explicit, fully realistic implementation of this mechanism, the Littlest 
Higgs model, is then discussed in some detail. Several other implementations, 
including simple group models and models with T parity, are also reviewed. 
Precision electroweak constraints on a variety of Little Higgs models are 
summarized. If a Little Higgs model is realized in nature, the predicted 
new particles should be observable at the Large Hadron Collider (LHC). The 
expected signatures, as well as the experimental sensitivities and the 
possible strategies for confirming
the Little Higgs origin of new particles, are discussed. 
Finally, several other related topics are briefly reviewed, including the
ultraviolet completions of Little Higgs models, as well as the 
implications of these models for flavor physics and cosmology.
\end{Abstract}
\vfill
\end{titlepage}
\tableofcontents
\newpage
\renewcommand{\thefootnote}{\arabic{footnote}}
\setcounter{footnote}{0}

\section{Introduction}

The standard model (SM) is a theory of electromagnetic, weak and strong interactions, whose predictions are in 
excellent agreement with the results of all particle physics experiments performed to date. Theorists, however, regard the SM as an effective theory, which is adequate at the presently explored energy scales but must become inadequate at a certain higher energy scale $\Lambda$. At the very least, the SM, which does not include gravity, must break down at the Planck energy scale $M_{\rm Pl}$ where the gravitational interactions become comparable in strength to other forces. More interestingly, there are serious theoretical reasons to believe that the SM breaks down much earlier, at the TeV scale. The arguments are based on the incompleteness of the SM description of electroweak symmetry breaking (EWSB). In the SM, this symmetry is assumed to be broken by the Higgs mechanism. 
The experimentally measured masses of $W$ and $Z$ bosons determine the vacuum expectation value (vev) of the Higgs field, $v\approx 250$ GeV, indicating that the Higgs mass parameter $\mu$ should be around the same scale. Moreover, precision electroweak data in the SM prefer a light Higgs boson: $m_h=\sqrt{2}\mu \lapproxeq 245$ GeV at 95\% c.l.~\cite{PDG}. In the SM, however, the parameter $\mu$ receives quadratically divergent one-loop radiative corrections. Assuming that the new physics at the scale $\Lambda$ cuts off the divergence gives an estimate
\beq
\delta \mu^2\sim g^2\Lambda^2/(16\pi^2), 
\eeq{badguy}
where $g$ is a gauge (or Yukawa) coupling constant. Barring the possibility of fine-tuning between the quantum corrections and the bare value of $\mu$, Eq.~\leqn{badguy} implies an upper bound on $\Lambda$ of approximately 2 TeV:  new physics must appear at or below this scale.
This generic prediction is particularly exciting today, since the Large Hadron Collider (LHC) will provide our first opportunity to explore the TeV energy scale experimentally in the near future.
 
Several theoretical extensions of the SM, attempting to provide a more satisfactory picture of EWSB and conjecture the structure of the theory at the TeV scale, have been proposed in the last three decades. 
Well-known examples include supersymmetric (SUSY) models, such as the minimal supersymmetric standard model (MSSM), and ``technicolor'' (TC) models which do not contain a Higgs boson, relying instead on strong dynamics to achieve EWSB. An intriguing alternative possibility is that a light Higgs boson exists, but is a composite particle, a bound state of more fundamental constituents held together by a new strong force~\cite{c1,c2,c3,c4,c5}. 
In this scenario, $\Lambda$ is the energy scale where the composite nature of the Higgs becomes important, which roughly coincides with the confinement scale of the new strong interactions. Unfortunately, precision electroweak data rule out new strong interactions at scales below about 10 TeV. To implement the composite Higgs without fine tuning, an additional mechanism is required to stabilize the ``little hierarchy'' between the Higgs mass and the strong interaction scale. 

In analogy with the pions of QCD, one can attempt to explain the lightness of the Higgs by interpreting it as a {\it Nambu-Goldstone boson} (NGB) corresponding to a spontaneously broken global symmetry of the new strongly interacting sector. However, gauge and Yukawa couplings of the Higgs, as well as its self-coupling, must violate the global symmetry explicitly, since an exact NGB only has derivative interactions. Quantum effects involving the symmetry-breaking interactions generate a potential, including a mass term, for the Higgs. Generically, this mass term is of the same size, Eq.~\leqn{badguy}, as in a model where no global symmetry exists to protect it: that is, the NGB nature of the Higgs is completely obliterated by quantum effects, and cannot be used to stabilize the little hierarchy. A solution to this difficulty has been proposed\footnote{Interestingly, the key insight behind this proposal, the collective symmetry breaking mechanism, was gleaned, via the ``dimensional deconstruction'' technique~\cite{decon,decon1}, from a study of five-dimensional models where the Higgs emerges as the fifth component of the 5D gauge field~\cite{A5}. Most Little Higgs theories reviewed in this article, however, do not have a simple five-dimensional or ``theory space'' interpretation.}
by Arkani-Hamed, Cohen and Georgi~\cite{bigmoose}. They argued that the gauge and Yukawa interactions of the Higgs can be incorporated in such a way that a quadratically divergent one-loop contribution to the Higgs mass is {\it not} generated. The cancellation of this contribution occurs as a consequence of the special ``collective'' pattern in which the gauge and Yukawa couplings break the global symmetries. The remaining quantum loop contributions to $\mu$ are much smaller, and no fine tuning is required to keep the Higgs sufficiently light if the strong coupling scale is of order 10 TeV: the little hierarchy is stabilized. 
``Little Higgs'' (LH) models incorporate the collective symmetry breaking mechanism to obtain natural and realistic theories of EWSB with a light composite Higgs boson. Many such models have been constructed in the last three years. 

All LH models contain new particles with masses around the 1 TeV scale. The interactions of these particles can be described within perturbation theory, and detailed predictions of their properties can be made. These states cancel the one-loop quadratically divergent contributions to the Higgs mass from SM loops. They provide distinct signatures that can be searched for at future colliders, as well as induce calculable, and often sizable, corrections to precision electroweak observables. At an energy scale of order 10 TeV, the LH description of physics becomes strongly coupled, and the LH model needs to be replaced by a more fundamental theory, its ``ultraviolet (UV) completion.'' The UV completion could be, for example, a QCD-like gauge theory with a confinement scale around 10 TeV. 

The goal of this article is to review the proposed LH models, the constraints placed on them by existing experimental data, and their predictions for future experiments\footnote{For another recent review of the Little Higgs models, see Ref.~\cite{review}.}.
The article is organized as follows: section~\ref{NGB} discusses a simple toy model attempting to realize the Higgs as an NGB. 
The toy model suffers from the ``little hierarchy'' problem; we then explain how the collective symmetry breaking mechanism, forming the backbone of the LH models, resolves this difficulty. We go on to present several fully realistic models implementing this mechanism: the ``Littlest Higgs'' model is discussed in detail in section~\ref{littlest}, while a number of other possibilities are reviewed in section~\ref{other}. Section~\ref{pew} will discuss the present constraints on the parameters of the LH models, dominated by the bounds from precision electroweak observables. Section~\ref{collide} covers the collider phenomenology of the LH models; the main focus is on the signatures that should be observed at the LHC if these models are realized in nature. Section~\ref{misc} provides a brief review of several other aspects of LH models studied in the literature, such as their possible ultraviolet completions, flavor physics, and cosmology.
Finally, section~\ref{conc} contains the conclusions.

\section{Higgs as a Nambu-Goldstone Boson: General Considerations}
\label{NGB}

To gain a better understanding of the issues involved in realizing the Higgs as a Nambu-Goldstone boson, in this section we will consider a simple toy model which incorporates this idea. We will discuss the phenomenological difficulties faced by the toy model, and describe the ``Little Higgs'' recipe for constructing a model that can avoid these difficulties~\cite{LH}. 

\subsection{A Toy Model}
\label{toy}

Consider a theory with a global $SU(3)$ symmetry, spontaneously broken to an $SU(2)$ subgroup, at a scale $f$, by a vacuum condensate $\Sigma_0$ transforming in the fundamental representation:
\beq
\Sigma_0 \,=\, \left(\begin{array}{c}
0 \\ 0\\ f\\ \end{array}\right).
\eeq{3vev}
According to Goldstone's theorem, the theory contains one massless field for each broken global symmetry generator; these fields are referred to as ``Nambu-Goldstone bosons'', or NGBs. Out of the 8 generators of $SU(3)$, 3 remain unbroken; thus, there are 5 NGBs in our theory, denoted by $\pi^a(x)$, $a=1\ldots 5$. These 5 fields decompose into a complex doublet and a singlet representation of the unbroken $SU(2)$ global symmetry; we will denote the doublet and the singlet by $H(x)$ and $s(x)$, respectively.  The dynamics of these fields below the ``cutoff'' scale $\Lambda\sim 4\pi f$ can be conveniently described in terms of an $SU(3)/SU(2)$ ``non-linear sigma model'' (nlsm)\footnote{For a pedagogical introduction to non-linear sigma models, see~\cite{Georgibook}.}, whose Lagrangian contains all possible Lorentz-invariant, local operators built out of the ``sigma field,''
\beq
\Sigma(x) = \frac{1}{f} \exp \left(\frac{2i \pi^a(x) X^a}{f}\right) \Sigma_0,
\eeq{sigma_def3}
and an arbitrary number of derivatives. Here the sum over $a=1\ldots 5$ is implicit, and the $X^a$ are broken $SU(3)$ generators. More explicitly,
\beq
2\pi^a(x) X^a \,=\, \left(\begin{array}{cc}
0 & H(x) \\ H^\dagger(x) & 0\\ \end{array}\right) 
\,+\,\frac{s(x)}{2\sqrt{2}}\left(\begin{array}{cc}
\iden & 0 \\ 0 & -2\\ \end{array}\right)\,,
\eeq{toy_pion}
where $\iden$ is a $2\times2$ unit matrix. Note that $\Sigma^\dagger \Sigma=1$; this constraint limits the number of independent operators than can be written at each order in the derivative expansion of the Lagrangian. The dynamics of the theory at low energies is determined by the terms with the smallest number of derivatives, while the higher-derivative terms give subdominant corrections. The leading term is given by
\beq
{\cal L}_{\rm kin} \,=\, f^2\, \partial_\mu \Sigma^\dagger \partial^\mu \Sigma,
\eeq{kinterm3}
where the normalization is chosen to yield canonically normalized kinetic terms for the NGB fields. 

To build a theory of electroweak symmetry breaking, we would like to identify a subset of the NGBs with the Higgs field of the SM. 
If the global $SU(3)$ symmetry were exact, the NGBs would only have derivative interactions; this is unacceptable since the SM Higgs must possess gauge interactions as well. The way around this difficulty is to introduce explicit breaking of the global symmetry by {\it gauging} an $SU(2)$ subgroup\footnote{For simplicity, we will omit the $U(1)$ component of the SM gauge structure in the discussion of this section; it will be reintroduced when experimental constraints and more realistic models are considered.} of the global $SU(3)$. Explicitly, this amounts to replacing the derivatives in Eq.~\leqn{kinterm3} by their covariant counterparts:
\beq
\partial_\mu \,\longrightarrow \,D_\mu \,=\, \partial_\mu - i g W_\mu^a(x) Q^a,
\eeq{covdiv3}
where $g$ is the gauge coupling constant, the $W_\mu^a$ are the $SU(2)$ gauge fields, and the $Q^a$ ($a=1\ldots 3$) are gauged generators given by
\beq
Q^a \,=\, \left(\begin{array}{cc}
\sigma^a/2 & 0 \\ 
0 & 0\\ 
\end{array}\right),
\eeq{gauged3}
the $\sigma^a$ being the standard Pauli matrices. After this replacement, Eq.~\leqn{kinterm3} contains the required $SU(2)$ gauge interactions for the field $H$, which can then be identified with the SM Higgs. 
Note that the Higgs field remains massless at tree level, since no explicit mass terms have been introduced. On the other hand, explicit  tree-level breaking of the global symmetry by the gauge interactions in Eq.~\leqn{covdiv3} will result in the Higgs acquiring a mass term, $\mu^2H^\dagger H$, as well as a quartic coupling, $\lambda (H^\dagger H)^2$, via quantum effects. In particular, if a negative $\mu^2$ and a positive $\lambda$ are generated, electroweak symmetry breaking will be triggered.  

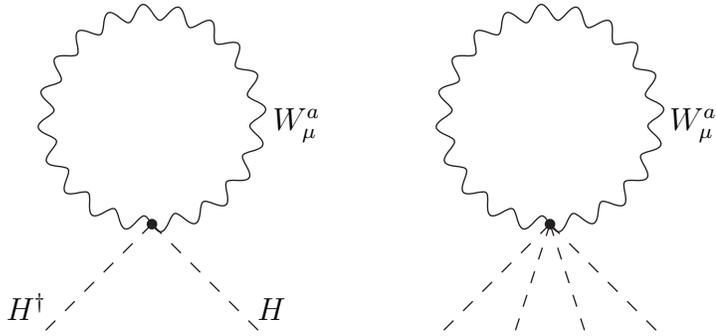
\begin{figure}[t]
\begin{center}
\begin{picture}(250,100)(0,0)
\DashLine(1,1)(41,41){5}
\DashLine(41,41)(81,1){5}
\Vertex(41,41){2}
\PhotonArc(41,81)(40,90,270){3}{10}
\PhotonArc(41,81)(40,270,360){3}{5}
\PhotonArc(41,81)(40,0,90){3}{5}
\Text(87,77)[bl]{$W^a_\mu$}
\Text(1,5)[br]{$H^\dagger$}
\Text(81,5)[bl]{$H$}
\DashLine(151,1)(191,41){5}
\DashLine(191,41)(231,1){5}
\DashLine(191,41)(178,1){5}
\DashLine(191,41)(204,1){5}
\Vertex(191,41){2}
\PhotonArc(191,81)(40,90,270){3}{10}
\PhotonArc(191,81)(40,270,360){3}{5}
\PhotonArc(191,81)(40,0,90){3}{5}
\Text(237,77)[bl]{$W^a_\mu$}
\end{picture}
\vskip2mm
\caption{The leading (quadratically divergent) one-loop contributions to the 
Higgs boson mass (left) and quartic coupling (right) from the 
gauge sector in the $SU(3)/SU(2)$ toy model.}
\label{fig:bowtie}
\end{center}
\end{figure}

Let us analyze the Higgs potential induced by quantum effects in more detail. The leading contributions to the Higgs mass and the quartic coupling arise from the ``bow tie'' one-loop diagrams\footnote{We assume that the calculations are performed in the Lorentz gauge, $\partial_\mu W^\mu=0$. In this gauge, the bow tie diagram provides the only contibution to the Higgs mass at one loop; the bubble diagram, involving the cubic Higgs gauge coupling, does not contribute.} shown in Fig.~\ref{fig:bowtie}. (The vertex in the diagram on the right appears when the $\Sigma$ fields in Eq.~\leqn{kinterm3} are expanded to cubic order in $H$.) Both diagrams are quadratically divergent, and the loop momentum integrals need to be cut off at a certain ultraviolet (UV) energy scale to obtain a finite result. The nlsm under consideration is an effective theory, valid below the cutoff energy scale $\Lambda\sim 4\pi f$. (At this scale, the tree-level NGB scattering amplitudes computed within the nlsm become inconsistent with unitarity, indicating that the nlsm description of physics breaks down and needs to be replaced with a more fundamental theory, the ``UV completion'' of the nlsm.) Assuming that the divergences in the bow tie diagrams are cut off at the same scale, we obtain an estimate
\beq
\mu^2 \,=\, c \frac{g^2}{16\pi^2}\,\Lambda^2 \sim c g^2 f^2,~~~
\lambda \,=\, c^\prime \frac{g^2}{f^2} \frac{1}{16\pi^2}\,\Lambda^2 \sim 
c^\prime g^2\,,
\eeq{mH3}
where $c$ and $c^\prime$ are order-one numbers whose exact values depend on the details of physics at the scale $\Lambda$ and can only be computed if the UV completion is specified. Assuming that $c$ is negative and $c^\prime$ positive, EWSB is induced at a scale of order $f$. After EWSB, three of the four components of the field $H$ are absorbed by the $W$ and $Z$ gauge bosons; the fourth component remains as a physical Higgs boson, with the mass $m_H=\sqrt{2}|\mu|\approx \sqrt{c} gf$. Precise measurements of the properties of $W$ and $Z$ bosons put an upper bound on the Higgs mass, $m_H \lapproxeq 245$ GeV at the 95\% c.l.~\cite{PDG}. Unless the parameter $c$ is numerically small, which would require fine-tuning, the scale $f$ has to be in the 200-300 GeV range, which in turn implies that the cutoff $\Lambda$ is, at most, around 3 TeV. Following the usual logic of Wilsonian effectve field theories, all possible non-renormalizable operators consistent with the symmetries of the low-energy theory should be assumed to be generated by quantum effects at the scale $\Lambda$, and they should be included in the Lagrangian with order-one coefficients. This, however, leads to phenomenological difficulties. For example, the Lagrangian will include the operators 
\beq
{\cal O}_1 \,=\,\frac{1}{\Lambda^2} \left| H^\dagger D_\mu H \right|^2,~~~{\cal O}_2 \,=\,\frac{1}{\Lambda^2} \left( H^\dagger \sigma^a H\right) W^a_{\mu\nu} B^{\mu\nu},
\eeq{ops3}
where $W^a_{\mu\nu}$ and $B_{\mu\nu}$ are the canonically normalized SM $SU(2)_L$ and $U(1)_Y$ gauge field strengths, respectively. Such terms are inconsistent with precision electroweak 
constraints for $\Lambda\lapproxeq 3$ TeV: the experimental upper bound on the coefficient of ${\cal O}_1$ is approximately $1/(5~{\rm TeV})^2$, while the bound for ${\cal O}_2$ is $\sim 1/(9~{\rm TeV})^2$~\cite{BPRS,HS}. Thus, in the absence of fine tuning, the simple realization of the Higgs as a pseudo-Goldstone boson considered here is ruled out by data. This is a manifestation of the ``little hierarchy'' problem, which is generic for theories postulating new strong dynamics as part of the EWSB mechanism: while the known scale of EWSB implies that new physics should appear at a scale of $\sim 1$ TeV, the precision electroweak constraints indicate that {\it generic, strongly coupled} new physics, parametrized by operators such as~\leqn{ops3}, cannot appear until $\sim 10$ TeV scale. 

\subsection{Collective Symmetry Breaking}

Can our toy model be modified to avoid the phenomenological difficulties we have found? A way to do this is suggested by the following observation: If the scale $f$ could be raised to, at least, about 1 TeV, the strong coupling scale $\Lambda$ would be of order $4\pi f\sim 10$ TeV, rendering the contributions of the strongly coupled UV physics to the precision electroweak observables sufficiently small to have escaped detection thus far. Since the Higgs mass parameter $\mu$ is forced to be at most of order 200 GeV by data, raising $f$ without fine tuning requires that the toy-model relation between these two scales, Eq.~\leqn{mH3}, be modified. Suppose that a model is found in which the one-loop quadratic divergence in the Higgs mass parameter is cut off at a scale $f$, rather than $\Lambda$. In such a model, the leading one-loop contribution to the Higgs mass parameter can be estimated as
\beq
\mu^2 \sim \frac{g^2}{16\pi^2}f^2 \log \frac{\Lambda^2}{f^2} \sim \frac{g^2}{8\pi^2} f^2 \log (4\pi),
\eeq{onelooplog}
which is of the right order of magnitude to generate the correct EWSB scale if $f\sim 1$ TeV {\it and} an order-one quartic coupling is generated. Note that the quadratic divergences at two- and higher-loop orders do {\it not} need to be cut off at $f$: the two-loop contribution with a cutoff at $\Lambda$ is given by
\beq
\delta \mu^2 \sim \frac{g^4}{(16\pi^2)^2}\,\Lambda^2 \sim \frac{g^4}{16\pi^2} f^2,
\eeq{twoloopquad}
and is subdominant to the log-enhanced one-loop contribution in~\leqn{onelooplog}. Thus, we would like to find extensions of our toy model in which the Higgs mass parameter is protected from one-loop quadratic divergences above the scale $f$. Little Higgs models achieve this goal in a simple and elegant way.

In Little Higgs models, the Higgs boson is embedded among the NGB fields arising when a global symmetry $G$ is broken to a subset $H$ at a scale $f$, assumed to be around a TeV. The NGBs are described by an $G/H$ non-linear sigma model. To describe the gauge interactions of the Higgs, a subgroup of $G$ must be weakly gauged. Unlike the toy model considered above, the gauged subgroup in Little Higgs models is not simple\footnote{The following describes the structure of the so-called ``product group'' Little Higgs models. There exists a second class of models, ``simple group'' models, in which the collective symmetry breaking mechanism is implemented in a slightly different way; see section~\ref{simple}.}, but is a direct product of two (or more) factors, $G_1\times G_2 \times \ldots$, each of which contains an $SU(2)\times U(1)$ subgroup. The gauged subgroup is embedded in $G$ in such a way that 
each of the $G_i$ factors commutes with a subgroup of $G$ that acts non-linearly on the Higgs. In other words, if only one of the $G_i$ factors is gauged, the unbroken global symmetry of the theory is sufficient to ensure that the Higgs is an {\it exact} Nambu-Goldstone boson, and is therefore massless to all orders in perturbation theory and even non-perturbatively. It is only when the full $G_1\times G_2 \times \ldots$ group is gauged that the Higgs ceases to be an exact NGB, and acquires non-derivative interactions. This structure is referred to as ``collective'' breaking of the global symmetries by gauge interactions. It implies that any non-vanishing quantum contribution to the Higgs mass parameter must necessarily be proportional to a product of {\it all} the gauge coupling constants corresponding to the different $G_i$ factors: setting any one of the coupling constants to zero must result in a vanishing contribution. On the other hand, the bow tie diagrams of the kind shown in Fig.~\ref{fig:bowtie}, which can induce the one-loop quadratic divergence in the Higgs mass parameter, only involve a single gauge coupling, corresponding to the gauge boson running in the loop. The collective symmetry breaking mechanism of the Little Higgs theories ensures that such diagrams are canceled. 

The extended gauge group $G_1\times G_2 \times \ldots$ of the LH models is typically broken down to the SM $SU(2)_L\times U(1)_Y$ at a scale $f$ by the same condensates that break $G\to H$. The models then contain additional gauge bosons at the TeV scale. In the mass eigenbasis, the vanishing of the one-loop quadratic divergence can be understood as a result of a cancellation between the SM bow tie diagrams and their counterparts involving the TeV-scale bosons. The relation between the couplings of these states to the Higgs is not accidental, but is enforced by the collective symmetry breaking mechanism. As we discuss specific Little Higgs models in the rest of this review, we will see explicit examples of how such cancellations work. 

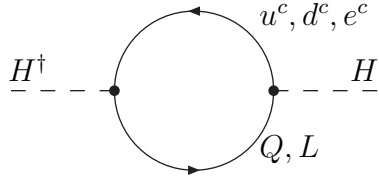
\begin{figure}[t]
\begin{center}
\begin{picture}(150,60)(0,0)
\DashLine(1,30)(40,30){5}
\DashLine(100,30)(139,30){5}
\Vertex(40,30){2}
\Vertex(100,30){2}
\ArrowArc(70,30)(30,0,180)
\ArrowArc(70,30)(30,180,360)
\Text(0,34)[bl]{$H^\dagger$}
\Text(140,34)[br]{$H$}
\Text(95,55)[bl]{$u^c, d^c, e^c$}
\Text(95,5)[bl]{$Q, L$}
\end{picture}
\vskip2mm
\caption{One-loop contribution to the Higgs boson mass from 
fermion loops.}
\label{fig:smtoploop}
\end{center}
\end{figure}

In addition to the gauge couplings of the Higgs, a realistic model needs to incorporate its Yukawa interactions. In a generic model with a cutoff $\Lambda$ and explicit breaking of the global symmetries by Yukawa interactions, the quadratically divergent one-loop diagram in Fig.~\ref{fig:smtoploop} induces a contribution to the Higgs mass parameter of order
\beq
\mu^2 \,=\, c_i \frac{y_i^2 N^c_i}{16\pi^2}\,\Lambda^2 \sim c_i y_i^2 N_i^c f^2,
\eeq{toptoy}
where $y_i$ is the Yukawa coupling of the fermion running in the loop, and $N^c_i=1$ for leptons and 3 for quarks. With the exception of the top quark, all SM fermions have small Yukawa couplings, $y_i \lapproxeq 0.03$, and their contribution does not induce any fine tuning in the Higgs mass for $\Lambda\sim 10$ TeV. The top quark, on the other hand, has a Yukawa of order one, and the quadratic divergence induced by top loops needs to be eliminated to avoid fine tuning. Again, this can be achieved if several Yukawa-like couplings are introduced in the Lagrangian, each one by itself preserving enough global symmetry to ensure exact vanishing of the Higgs mass. Quantum corrections to the Higgs mass must involve {\it all} Yukawa couplings, and no quadratically divergent diagrams with this property can be drawn. Explicit examples of models implementing the collective symmetry breaking mechanism in the top sector will be discussed below.

\section{The Littlest Higgs Model}
\label{littlest}

A number of fully realistic Little Higgs models of EWSB, based on the collective symmetry breaking mechanism outlined above, have been constructed. The ``Littlest Higgs'' model, proposed by Arkani-Hamed, Cohen, Katz and Nelson in Ref.~\cite{LH}, is one of the most economical and attractive implementations. Most of the phenomenological studies up to date have been performed in the context of this model or its modifications. This section contains a detailed review of the Littlest Higgs model.

\subsection{Gauge and Scalar Sector}

The Littlest Higgs (L$^2$H) model~\cite{LH} embeds the electroweak sector of the standard model in an $SU(5)/SO(5)$ non-linear sigma model.  Consider a theory with a global $SU(5)$ symmetry,
spontaneously broken to an $SO(5)$ subgroup at an energy scale $f\sim 1$ TeV by a vacuum condensate transforming in the symmetric tensor representation of $SU(5)$. It is convenient to choose a basis in which the condensate is proportional to  
\beq
\Sigma_0 \,=\, \left(\begin{array}{ccc}
0& 0& \iden\\
0& 1& 0\\
\iden& 0& 0\\
\end{array}\right), 
\eeq{sigma0}
where $\iden$ represents a unit $2\times 2$ matrix. As in the toy model of section~\ref{toy}, the dynamics of the theory at low energies can be completely described in terms of the Nambu-Goldstone degrees of freedom, which are massless. As usual, there is one Goldstone particle for each broken generator $X^a$; for the $SU(5)\to SO(5)$ breaking, there are $24-10=14$ broken generators, and thus 14 NGB fields, which we will denote by $\pi^a(x)$. The interactions of the NGBs at energy scales below $\Lambda\sim 4\pi f$ are described by an $SU(5)/SO(5)$ non-linear sigma model, whose Lagrangian contains all possible Lorentz-invariant, local operators built out of the field
\beq
\Sigma(x) \,=\, e^{i \Pi/f} \Sigma_0 e^{i \Pi^T/f} \,=\, e^{2i \Pi/f} \Sigma_0,
\eeq{sigma_def}
and an arbitrary number of derivatives. Here we defined the ``pion matrix''
\beq
\Pi(x)=\sum_{a=1}^{14} \pi^a(x) X^a, 
\eeq{pionmatrix}
and used the relation $X^a\Sigma_0=\Sigma_0 X^{aT}$, obeyed by the broken generators, in the last step.  Just as in the toy model, the relation $\Sigma^\dagger \Sigma=1$ greatly reduces the number of independent operators than can be written at each order in the derivative expansion. 

The Higgs field of the SM is identified with a subset of the NGB degrees of freedom of the theory. To describe the gauge interactions of the Higgs, the global symmetry is explicitly broken by gauging an $[SU(2)\times U(1)]^2$ subgroup of the $SU(5)$. The gauged generators have the form
\begin{eqnarray}\label{gauged}
&Q_1^a=\left( \begin{array}{ccc} \sigma^a/2 &0 & 0 \\
0 & 0 & 0\\ 0 & 0 & 0
\end{array}\right), \ \ \ &Y_1=
{\rm diag}(3,3,-2,-2,-2)/10\,,\nonumber \\
&Q_2^a=\left( \begin{array}{ccc} 0 & 0 & 0\\
0 & 0 & 0 \\
0 &0&-\sigma^{a*}/2\end{array} \right), & Y_2={\rm
diag}(2,2,2,-3,-3)/10~.
\end{eqnarray}
The Lagrangian of the gauged theory is obtained from the original nlsm by the following replacement:
\beq
\partial_\mu \Sigma \rightarrow D_\mu \Sigma =
\partial_\mu \Sigma - i \sum_{j=1}^2 \left[ g_j W_{j\mu}^a (Q_j^a \Sigma + \Sigma Q_j^{aT} )+ g'_j B_{j\mu}( Y_j \Sigma + \Sigma Y_j)\right]\,.
\eeq{cov_littlest}
Here, $B_j$ and $W^a_j$ are the $U(1)_j$ and
$SU(2)_j$ gauge fields, respectively, and $g^\prime_j$ and $g_j$ are the corresponding coupling constants. For example, the kinetic term for the $\Sigma$ field, which has the lowest number of derivatives (two) among the allowed non-trivial Lagrangian terms and therefore dominates the low energy dynamics, has the form
\beq
{\cal L}_{\rm kin} \,=\, \frac{f^2}{8} {\mathrm Tr} (D_\mu \Sigma) (D^\mu 
\Sigma)^\dagger.
\eeq{kinL}
Here, the normalization has been chosen to ensure that the fields $\pi^a$ 
have canonically normalized kinetic terms (the generators are normalized
according to $\Tr(X^aX^b)=\delta^{ab}$). 
Note that no explicit mass terms have been introduced for the NGBs, which remain massless at tree level. On the other hand, tree-level breaking of the global symmetry by the gauge interactions in Eq.~\leqn{kinL} will result in the NGBs acquiring mass terms via quantum effects.

The gauge symmetry breaking in the L$^2$H model occurs in two stages: first, the condensate $\Sigma_0$ breaks the extended gauge group $\left[ SU(2) \times U(1) \right]^2$ down to the diagonal subgroup, which is identified with the standard model electroweak group $SU(2)_L \times U(1)_Y$. This is a tree level effect in the nlsm, occuring at the scale $f\sim 1$ TeV. Then, at the scale $v=246$ GeV, the usual EWSB occurs, breaking $SU(2)_L \times U(1)_Y\to U(1)_{\rm em}$. EWSB is triggered by a radiatively induced Higgs potential. Let us consider the first stage of the symmetry breaking, $\left[ SU(2) \times U(1) \right]^2 \to SU(2)_L \times U(1)_Y$. (The EWSB stage will be considered in section~\ref{ewsb}.) The gauge couplings of the unbroken diagonal subgroup are given by
\beq
g \,=\,\frac{g_1g_2}{\sqrt{g_1^2+g_2^2}},~~~g^\prime \,=\,\frac{g^\prime_1g^\prime_2}{\sqrt{g_1^{\prime 2}+g_2^{\prime 2}}};
\eeq{ewcouple}
these are set equal to the SM weak and hypercharge gauge couplings, respectively. This identification leaves two free dimensionless parameters in this sector of the theory; it is convenient to use the two mixing angles, $\psi$ and $\psi^\prime$, defined by
\beq
\tan \psi = {g_2 \over g_1},~~~\tan \psip = {g_2^\prime \over g_1^\prime}.
\eeq{psis}
The linear combinations of the gauge fields $W^a_j$, $B_j$ that acquire TeV-scale masses are given by
\beq
W_H^a \,=\, -\cos \psi \,W_1^a + \sin \psi \,W_2^a,~~
B_H \,=\, -\cos\psip \,B_1 + \sin\psip \,B_2,
\eeq{massive}
and their masses are
\beq
M(W_H) \,=\, \frac{g}{\sin 2\psi}\,f\,,~~~
M(B_H) \,=\, \frac{g^\prime}{\sqrt{5} \sin 2\psip}\,f\,.
\eeq{WHmasses}
The orthogonal linear combinations, 
\beq
W_L^a \,=\, \sin \psi \,W_1^a + \cos \psi \,W_2^a,~~
B_L \,=\, \sin\psip \,B_1 + \cos\psip \,B_2,
\eeq{massless}
remain massless at this stage. 

The fourteen NGBs of the $SU(5)/SO(5)$ breaking 
decompose into representations of the electroweak gauge group as follows:
\beq
\mathbf{1_0} \oplus \mathbf{3_0} \oplus \mathbf{2_{1/2}}
\oplus \mathbf{3_{1}},
\eeq{reps}
where the subscripts indicate the hypercharges.
Let us denote the fields in these four representations by
$\eta$, $\omega$, $H$ and $\phi$, respectively. The field $H$ has the 
appropriate quantum numbers to be identified with the SM Higgs.
Explicitly, the pion matrix in terms of these fields has the form
\beq
\Pi\,= \left(\begin{array}{ccccc}
-\frac{\omega^{0}}{2}- \frac{\eta}{\sqrt{20}} & -\omega^+/\sqrt{2} & H^+/\sqrt{2} & -i 
\phi^{++} & -i
\frac{\phi^{+}}{\sqrt{2}} \\
-\omega^-/\sqrt{2} & \omega^0/2- \eta/\sqrt{20} & H^0/\sqrt{2}& -i
\frac{\phi^{+}}{\sqrt{2}} & \frac{-i \phi^0 +\phi_P^0}{\sqrt{2}} \\
H^-/\sqrt{2} &  H^{0*}/\sqrt{2}& \sqrt{4/5} \eta  & H^+/\sqrt{2} & H^0/\sqrt{2} \\
i \phi^{--} & i \frac{\phi^{-}}{\sqrt{2}} & H^-/\sqrt{2} &
-\omega^0/2 - \eta/\sqrt{20} &
- \omega^-/\sqrt{2} \\
i \frac{\phi^{-}}{\sqrt{2}} & \frac{i \phi^0 +\phi_P^0}{\sqrt{2}} &
H^{0*}/\sqrt{2} & - \omega^+/\sqrt{2} &\frac{\omega^0}{2}- \frac{\eta}{\sqrt{20}}
\end{array}\right),
\eeq{pions}
where the superscripts indicate the electric charge. (The normalizations are chosen so that all fields are canonically normalized.) The fields $\eta$ and $\omega$ are absorbed by the $B_H$ and $W^a_H$ heavy gauge bosons, respectively, while $H$ and $\phi$ remain physical (and massless) at this stage. 

When quantum corrections involving gauge interactions are included, the $H$ and $\phi$ fields are no longer exact NGBs: they acquire a potential, which can be computed using the standard Coleman-Weinberg approach~\cite{CW}, see section~\ref{ewsb}. Before proceeding with the full calculation, however, it is useful to understand the crucial feature of this model: the cancellation of quadratically divergent contributions to the Higgs boson mass at one loop. 

We start by observing that the gauge generators are embedded in the $SU(5)$ in such a way that any given generator commutes with an $SU(3)$ subgroup of the $SU(5)$. Indeed, the generators $Q_1^a$ and $Y_1$ commute with the $SU(3)$ generators embedded in the lower-right corner of the $SU(5)$ matrices, whereas $Q_2^a$ and $Y_2$ commute with the ``upper-left corner'' $SU(3)$. This implies that if one pair of gauge couplings ($g_1, g_1^\prime$ or $g_2, g_2^\prime$) is set to zero, 
the model possesses an exact $SU(3)$ global symmetry, spontaneously broken down to an $SU(2)$ subgroup by the vev $\Sigma_0$. In both cases, the Higgs field $H$ is the Nambu-Goldstone boson corresponding to this breaking, and therefore is exactly massless at all orders in perturbation theory and even non-perturbatively (at least as long as gravitational effects are neglected). Thus, any diagram renormalizing the Higgs mass vanishes unless it involves at least {\it two} of the gauge couplings. In a generic theory, the one-loop quadratic divergence is due to the bow tie diagrams of the kind shown in Fig.~\ref{fig:bowtie}. Each of them involves a single gauge field, and therefore cannot contain more than a single gauge coupling. Thus, these diagrams are either absent in the 
L$^2$H theory, or their contributions cancel out; in either case, there is no one-loop quadratic divergence in the Higgs mass. Note that this argument does not hold for the triplet scalar field $\phi$, which acquires a TeV-scale mass 
from quadratically divergent one-loop diagrams.

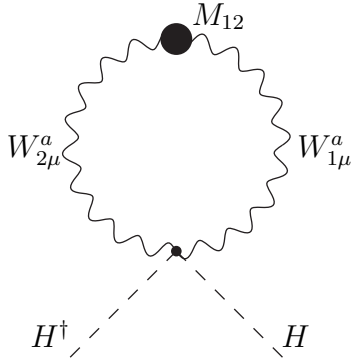
\begin{figure}[t]
\begin{center}
\begin{picture}(100,100)(0,0)
\DashLine(1,1)(41,41){5}
\DashLine(41,41)(81,1){5}
\Vertex(41,41){2}
\Vertex(41,121){6}
\Text(47,127)[bl]{$M_{12}$}
\PhotonArc(41,81)(40,90,270){3}{10}
\PhotonArc(41,81)(40,270,360){3}{5}
\PhotonArc(41,81)(40,0,90){3}{5}
\Text(87,77)[bl]{$W^a_{1\mu}$}
\Text(-2,77)[br]{$W^a_{2\mu}$}
\Text(1,5)[br]{$H^\dagger$}
\Text(81,5)[bl]{$H$}
\end{picture}
\vskip2mm
\caption{One-loop contribution to the Higgs boson (mass)$^{2}$ from the 
$SU(2)$ gauge sector in the Littlest Higgs model, in the gauge eigenbasis. 
The blob indicates an off-diagonal mass insertion.}
\label{fig:bowtiemass}
\end{center}
\end{figure}

A more explicit analysis of the cancellation~\cite{BPP}  shows that the bow-tie diagrams of the dangerous type are completely absent in the L$^2$H model. All quartic couplings involving two Higgs fields and two gauge bosons that appear in the model can be worked out from Eq.~\leqn{kinL}; they have the form
\beq
\frac{1}{4}\,H^\dagger H \,\left( g_1 g_2 \,W_1^{\mu a} \,W_{2\mu}^a + 
g_1^\prime g_2^\prime \,B_1^\mu \,B_{2\mu}\right)\,.
\eeq{higgs_coupl}
In a generic theory, one would also expect diagonal couplings such as
$g_1^2 W_1^2 H^\dagger H$ and $g_2^2 W_2^2 H^\dagger H$. These couplings would lead to quadratically divergent bow tie diagrams. The absence of the diagonal couplings in Eq.~\leqn{higgs_coupl}, which is a consequence of the collective symmetry breaking structure of the theory, guarantees the absence of such divergences. The only bow tie diagrams in the L$^2$H model include a mass insertion which mixes $W_1$ and $W_2$ (or $B_1$ and $B_2$) fields, see Fig.~\ref{fig:bowtiemass}. These diagrams diverge only logarithmically. 

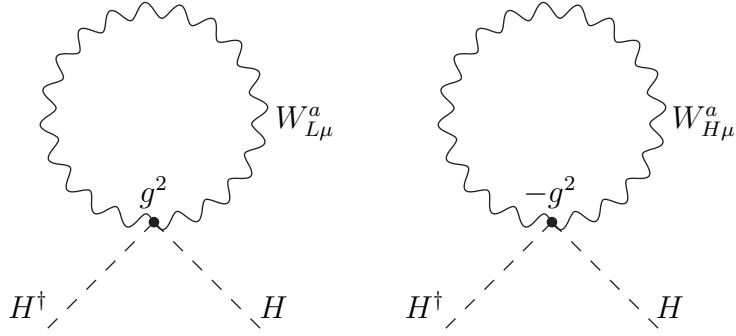
\begin{figure}[t]
\begin{center}
\begin{picture}(250,100)(0,0)
\DashLine(1,1)(41,41){5}
\DashLine(41,41)(81,1){5}
\Vertex(41,41){2}
\PhotonArc(41,81)(40,90,270){3}{10}
\PhotonArc(41,81)(40,270,360){3}{5}
\PhotonArc(41,81)(40,0,90){3}{5}
\Text(87,77)[bl]{$W^a_{L\mu}$}
\Text(1,5)[br]{$H^\dagger$}
\Text(81,5)[bl]{$H$}
\Text(41,49)[bc]{$g^2$}
\DashLine(151,1)(191,41){5}
\DashLine(191,41)(231,1){5}
\Vertex(191,41){2}
\PhotonArc(191,81)(40,90,270){3}{10}
\PhotonArc(191,81)(40,270,360){3}{5}
\PhotonArc(191,81)(40,0,90){3}{5}
\Text(237,77)[bl]{$W^a_{H\mu}$}
\Text(151,5)[br]{$H^\dagger$}
\Text(231,5)[bl]{$H$}
\Text(191,49)[bc]{$-g^2$}
\end{picture}
\vskip2mm
\caption{One-loop contributions to the Higgs boson (mass)$^{2}$ from the 
$SU(2)$ gauge sector in the Littlest Higgs model, in the mass eigenbasis.}
\label{fig:2bowties}
\end{center}
\end{figure}

The cancellation can also be understood in terms of the gauge boson mass eigenstates $W_{L/H}$ and $B_{L/H}$. In this basis, Eq.~\leqn{higgs_coupl} 
becomes~\cite{BPP}
\beqa 
& &\frac{1}{4}\,H^\dagger\,H \Bigl(g^2(W_{L\mu}^aW^{\mu a}_L - W_{H\mu}^aW^{\mu a}_H
- 2\cot 2\psi\, W_{H\mu}^aW^{\mu a}_L)
+ \CR & &~~~g^{\prime 2}(B_{L\mu}B^\mu_L - B_{H\mu}B^\mu_H
- 2\cot 2\psip\,B_{H\mu}B^\mu_L)\Bigr)\,. 
\eeqa{higgs_coupl1}
The diagonal couplings of the Higgs to the gauge bosons are now present, but the couplings to the light and heavy states have equal magnitude and opposite sign, ensuring the cancellation of quadratic divergences. This is illustrated in Fig.~\ref{fig:2bowties}. Again, the relation between the couplings is not an accident, but an unavoidable consequence of the symmetry structure of the theory. 

\subsection{Fermion Sector}
\label{fermions}

\begin{figure}[t]
\begin{center}
\includegraphics[width=15cm]{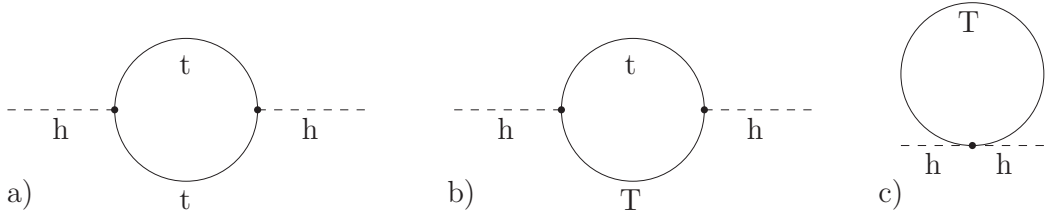}
\vskip2mm
\caption{One-loop contributions to the Higgs boson (mass)$^{2}$ from the top sector of the Littlest Higgs model.}
\label{fig:toploop}
\end{center}
\end{figure}

The largest quadratically divergent one-loop contribution to the Higgs mass parameter in the SM actually comes not from the gauge sector, but from the top quark loop shown in Fig.~\ref{fig:toploop} (a). To cancel this divergence, the top Yukawa coupling has to be extended to incorporate the collective symmetry breaking mechanism. In the L$^2$H model, this is achieved by introducing a pair of weak-singlet Weyl fermions $U_L$ and $U_R$ with electric charge $+2/3$. These are coupled to the third generation quark doublet $q_{3L}=(u_L, b_L)^T$ and the singlet $u_{3R}$ in the following way:
\beq
 {\cal L}_{\rm top} =   - {\lambda_1\over 2}f\, \chi_{L i}^\dagger   \epsilon_{ijk} \epsilon_{mn} \Sigma_{jm} \Sigma_{kn} u_{3R} - \lambda_2 f\, U^\dagger_L U_R + {\rm h.c.}\  ,
\eeq{top_yuk}
where 
\beq
\chi_L=\left(\begin{array}{c}\sigma_2 q_{3L} \\ U_L \end{array}\right)
\eeq{ubu}
is the ``royal'' $SU(3)$ triplet~\cite{Jarry}, and $\Sigma_{jm}$ denotes the $3\times 2$ upper right hand block of the $\Sigma$ field defined in Eq.~\leqn{sigma_def}. (The indices $i,j,k$ run between $1$ and $3$, and $m, n=4,5$.) The spectrum and interactions of the top quark and its partners can be obtained by expanding the $\Sigma$ fields in this Lagrangian to the desired order. Neglecting the EWSB effects, the mass eigenstates are given by
\beqa
  t_L = u_L,       &\qquad&   t_R =  {\lambda_2 u_{3R}-  \lambda_1 U_R\over
                      \sqrt{\lambda_1^2 + \lambda_2^2}}, \CR
            T_L = U_L,      & \qquad&  T_R = {\lambda_1 u_{3R} + \lambda_2 U_R\over
                                   \sqrt{\lambda_1^2 + \lambda_2^2}},
\eeqa{topmix}
with $t$ massless at this level and 
\beq
M_{T}= \sqrt{\lambda_{1}^{2} +\lambda_{2}^{2}} \, f.
\eeq{heavytopmass}
The couplings of these states to the Higgs, up to the quadratic order in $H$, have the form
\beqa
& & \lambda_1 \left( \sqrt{2} q_L^\dagger \tilde{H} - \frac{1}{f}  H^\dagger H U_L^\dagger\right) u_{3R}~+{~\rm h.c.} \CR & & =
\lambda_t  q_L^\dagger \tilde{H} t_R \,+\, \lambda_T q_L^\dagger \tilde{H} T_R - \frac{1}{\sqrt{2}f} (H^\dagger H) T_L^\dagger \left( \lambda_T T_R +\lambda_t t_R\right)~+{~\rm h.c.}
\eeqa{tophiggs}
where $\tilde{H}=i\sigma^2 H$, and we define
\beq
\lambda_t = \frac{\sqrt{2}\lambda_1\lambda_2}{\sqrt{\lambda_1^2+\lambda_2^2}},~~~~\lambda_T=\frac{\sqrt{2}\lambda_1^2}{\sqrt{\lambda_1^2+\lambda_2^2}}.
\eeq{yukawas}
The first term in the second line of Eq.~\leqn{tophiggs} is the usual SM up-type Yukawa coupling; after EWSB, the Higgs acquires a vev, and this coupling provides the $t$ with its mass, $m_t\approx\lambda_t v/\sqrt{2}$. The Higgs vev also generates corrections of order $v/f$ to Eqs.~\leqn{topmix} and~\leqn{heavytopmass}. Precise formulas for the
masses and mixing angles in this sector, valid to all orders in $v/f$, can be
found in Refs.~\cite{PPP,HMNP}.

It is easy to see that the Lagrangian of Eq.~\leqn{top_yuk} preserves the collective symmetry breaking pattern that was exhibited by the gauge sector of the model. If the coupling $\lambda_1$ is set to zero, the Higgs is completely decoupled from the top sector, whereas if $\lambda_2=0$, the upper-left corner global $SU(3)$ is unbroken (with $\chi_L$ transforming in the fundamental representation of this group) and the Higgs is an exact NGB. In both cases, the Higgs remains massless to all orders in perturbation theory and beyond. Any contribution to the Higgs mass must therefore involve both $\lambda_1$ and $\lambda_2$. Again, one-loop diagrams with this property are at most logarithmically divergent. This is most easily seen in the original basis, where the mass matrix is not diagonal. In this basis, all Higgs couplings only involve $\lambda_1$, with the  $\lambda_2$ appearing only in the masses; therefore, any non-vanishing contribution to the Higgs mass must contain mass insertions in the propagators, which reduces the maximal degree of divergence at one loop from quadratic to logarithmic.

In the mass eigenbasis, the one-loop contribution to the Higgs mass from the top sector involves the three diagrams shown in Fig.~\ref{fig:toploop}. The values of the diagrams are~\cite{PPP}
\beqa
     \mbox{a)} &=&  -6 \lambda_t^2 \int {d^4k\over (2\pi)^4}  {1\over k^2},\CR
     \mbox{b)} &=&  -6 \lambda_T^2 \int {d^4k\over (2\pi)^4}
                                           {1\over k^2- M_T^2},\CR \CR
     \mbox{c)} &=& +6 {\sqrt{2} \lambda_T\over f}
            \int {d^4k\over (2\pi)^4}  {M_T\over k^2 - M_T^2},
\eeqa{tloops} 
The quadratic divergences neatly cancel. It is interesting to note that this cancellation hinges on the relation
\footnote{Note that this relation differs, by a factor of $\sqrt{2}$, from that obtained in Ref.~\cite{PPP}. This is a consequence of a different normalization for $f$ used in that reference.}
\beq
  \frac{M_T}{f} =    { \lambda_t^2  + \lambda_T^2\over  \sqrt{2} \lambda_T}\ ,
\eeq{yuk_rel}
which could in principle be tested by the LHC experiments, see Section~\ref{collide} and Ref.~\cite{PPP}.

For fermions other than the top quark, the quadratically divergent diagrams in Fig.~\ref{fig:smtoploop} do not necessitate fine-tuning in the Higgs potential if the cutoff is at 10 TeV, due to the small values of the corresponding Yukawa couplings.  
Because of this, there is no need to implement collective symmetry breaking in this sector. Yukawa couplings for the light quarks of the up type can be generated by operators analogous to Eq.~\leqn{top_yuk}, but without the need for extra singlet states $U_{L,R}$. Yukawa couplings for the down-type quarks of all three generations and the charged leptons can be generated by the same operators with $\Sigma\to\Sigma^*$. This, however, is not the only possibility: for example, the up-type Yukawas can in general arise from terms of the form~\cite{CC2}
\beq
{\cal L} \sim Q^\dagger_L H X^r Y^s u_R \,+\, {\rm h.c.}, 
\eeq{Yuk_general}
where $Q_L$ and $u_R$ are the SM quark doublet and up-type singlet, respectively, $X$ and $Y$ are components of the $\Sigma$ field that get unit vevs, and $r$ and $s$ are positive integers. The charges of light fermions under the $[SU(2)\times U(1)]^2$ gauge group are constrained by the requirements that the charges under the diagonal subgroup coincide with the SM assignments, and that the Yukawa couplings be invariant. This implies that the weak doublets of the SM transform either as ({\bf 2,1}) or ({\bf 1,2}) under the two $SU(2)$ factors, whereas the weak singlets transform as ({\bf 1,1}). The $U(1)\times U(1)$ charges of each fermion can be written as $(RY, (1-R)Y)$, where $Y$ is its SM hypercharge and $R$ is a constant which depends on the precise form of the operator which generates the Yukawa coupling. Assuming that the $U(1)$ charges are universal within each generation of fermions, and the Yukawa couplings have the form~\leqn{Yuk_general}, Ref.~\cite{CC2} found that $R$ has to be quantized in units of 1/5. (Phenomenological studies frequently assume $R=0$ or $R=1$.) Note that there is no additional constraint on the fermion charges from anomaly considerations, since there may be additional fermions at the cutoff which cancel the anomalies involving the broken subgroup. The only constraint on the low-energy theory is the cancellation of anomalies involving the SM $SU(2)_L\times U(1)_Y$ gauge interactions, since the fermions in a chiral representation of this group can only have weak-scale masses. This requirement is however automatically satisfied: by construction, the light fermions have their usual SM charges under this group. 

\subsection{Scalar Potential and Electroweak Symmetry Breaking}
\label{ewsb}

At tree level, the Higgs field $H$ and the weak triplet scalar field $\phi$ have no non-derivative interactions. However, due to the explicit breaking of the global $SU(5)$ symmetry by gauge and Yukawa interactions, quantum corrections induce a Coleman-Weinberg (CW) potential~\cite{CW} for both $\phi$ and $H$, triggering electroweak symmetry breaking. The leading one-loop contribution to the CW
potential from quadratically divergent gauge loops is~\cite{LH}
\beq
V^{(quad)}_{\rm g}\,=\,\frac{\Lambda^2}{16\pi^2} {\rm Tr}\,M_V^2(\Sigma),
\eeq{CW_quad_gauge}
where $M_V^2(\Sigma)$ is the gauge boson mass matrix in an arbitrary $\Sigma$ background. Using Eq.~\leqn{kinL} and $\Lambda\sim 4\pi f$ yields
\beq
V^{(quad)}_{\rm g}\,=\,a f^4 \sum_j \left( g_j^2 \sum_b \Tr [Q_j^b \Sigma Q_j^{b*} \Sigma^*] \,+\, g_j^{\prime 2} \Tr [Y_j \Sigma Y_j^* \Sigma^*]\right),
\eeq{CWg1}
where $a$ is an order-one number whose precise value depends on the physics at the scale $\Lambda$, i.e. the UV completion of the L$^2$H theory. Expanding the $\Sigma$ fields up to quadratic order in $\phi$ and quartic order in $H$ yields
\beqa
V^{(quad)}_{\rm g} &=& a (g_1^2+g_1^{\prime 2}) f^2 \left|\phi_{ij} + \frac{i}{4f} (H_iH_j + H_j H_i) \right|^2 \CR & &+\,a (g_2^2+g_2^{\prime 2}) f^2 \left|\phi_{ij} - \frac{i}{4f} (H_iH_j + H_j H_i) \right|^2,
\eeqa{CWg2} 
where the sum over $i,j=1,2$ is implicit, and $H_i\equiv \sqrt{2} \Pi_{i3}$, $\phi_{ij}\equiv \Pi_{i,3+j}$. The only other quadratically divergent contribution to the one-loop CW potential is generated by top loops\footnote{The light fermion contributions can be neglected due to their small Yukawa couplings.}; it is given by
\beq
V^{(quad)}_{\rm t}\,=\,-a^\prime \lambda_1^2 \eps^{wx}\eps_{yz} \eps^{ijk} \eps_{kmn} \Sigma_{iw} \Sigma_{jx} \Sigma^{*my} \Sigma^{*nz} \,+\, {\rm h.c.},
\eeq{CWf0}
where $i,j,k,m,n=1\ldots 3$ and $w,x,y,z=4,5$. Expanding in terms of $H$ and 
$\phi$, we obtain
\beq
V^{(quad)}_{\rm t}\,=\,-a^\prime \lambda_1^2 f^2 \left|\phi_{ij} + \frac{i}{4f} (H_iH_j + H_j H_i) \right|^2.
\eeq{CWf1}
Note that the potentials~\leqn{CWg2},~\leqn{CWf1} do not contain a mass term for the Higgs field; as discussed above, this is a consequence of the collective symmetry breaking pattern, which prohibits the quadratically divergent contributions to the Higgs mass at one loop. However, a quadratically divergent mass term for $\phi$ is present, giving it a mass of order $f$:
\beq
M_\phi^2 = \left( a (g_1^2+g_1^{\prime 2}+g_2^2+g_2^{\prime 2}) - 
a^\prime \lambda_1^2 \right) f^2.
\eeq{phimass}
Successful EWSB is only possible for points in the parameter space where $M_\phi^2 > 0$, since otherwise electroweak symmetry is broken by a triplet vev of order $f$. At energies beneath the triplet mass, this field can be integrated out, resulting in a quartic potential for the Higgs, $\lambda (H^\dagger H)^2$, where
\beq
\lambda = a\,\frac{(g_1^2+g_1^{\prime 2}-a^\prime\lambda_1^2/a)(g_2^2+g_2^{\prime 2})}{g_1^2+g_1^{\prime 2}+g_2^2+g_2^{\prime 2}-a^\prime\lambda_1^2/a}.
\eeq{Hquart}
Thus, quadratically divergent one-loop diagrams generate an unsuppressed quartic coupling, as required to satisfy the lower bound on the Higgs boson mass, $m_h =\sqrt{\lambda/2} v \gapproxeq 114$ GeV~\cite{PDG}. 

A mass term for the Higgs field is generated at the one loop level by logarithmically divergent diagrams. The logarithmically enhanced contribution of the vector bosons to the one-loop CW potential is given by
\beq  
V_{\rm g}^{(log)} \,=\, \frac{3}{64\pi^2}\,\Tr M_V^4(\Sigma)\,\log \frac{M^2_V(\Sigma)}{\Lambda^2}.
\eeq{CW_gauge_log}
Expanding the sigma fields to quadratic order in $H$ yields a positive contribution to the Higgs mass parameter:
\beq
\mu_{\rm g}^2(H) = \frac{3}{64\pi^2}\left( 3g^2 M^2_{W_H} \log \frac{\Lambda^2}{M^2_{W_H}} \,+\, g^{\prime 2} M^2_{B_H} \log
\frac{\Lambda^2}{M^2_{B_H}}\right).
\eeq{mH_g}
The top loop contribution has the form
\beq
V_{\rm t}^{(log)} \,=\, -\frac{3}{16\pi^2}\,\Tr\left(M_t(\Sigma) M_t^\dagger(\Sigma) \right)^2\,\log \frac{M_t(\Sigma) M_t^\dagger(\Sigma)}{\Lambda^2},
\eeq{CW_top_log}
where $M_t(\Sigma)$ is the top mass matrix in an arbitrary $\Sigma$ background. This potential gives a {\it negative} contribution to the Higgs mass parameter:
\beq
\mu_{\rm t}^2(H) \,=\,- {3 \lambda_t^2 M_{T}^{2} \over 8\pi^2 }  
                   \log {\Lambda^2\over M_T^2} \ .
\eeq{mH_t}
Finally, the logarithmically enhanced part of the CW potential also receives a contribution from the scalar sector; the corresponding correction to the Higgs mass parameter is given by
\beq
\mu_{\rm s}^2(H) \,=\, {\lambda \over 16\pi^2 } M_\phi^2  
                   \log {\Lambda^2\over M_\phi^2} \ .
\eeq{mH_s}
Due to the large value of the top Yukawa coupling, the top loop contribution to the Higgs mass~\leqn{mH_t} typically dominates over the gauge and scalar contributions, triggering EWSB. The subleading contributions to the Higgs mass parameter include the finite one-loop terms, of order $f^2/(16\pi^2)$, and the quadratically divergent two-loop terms, of order $\Lambda^2/(4\pi)^4 \sim f^2/(16\pi^2)$. The precise values of these terms depend on the physics at scale $\Lambda$; however, both are parametrically smaller than the terms in Eqs.~\leqn{mH_g},~\leqn{mH_t} and~\leqn{mH_s} by a factor of $\log(4\pi)^2 \sim 5$. Thus, the LH model provides a robust mechanism of radiative EWSB via top loops.

After EWSB, the Higgs field can be decomposed as 
\beq
H\,=\,\left(\pi^+, \frac{v+h+i\pi^0}{\sqrt{2}} \right)^T\,,
\eeq{higgsische}
where $v=246$ GeV is the EWSB scale and $h$ is the physical Higgs field. The $\pi^\pm$ and $\pi^0$ fields are absorbed by the SM $W^\pm$ and $Z$ bosons\footnote{More precisely, the fields absorbed by the $W^\pm$ and $Z$ bosons are {\it predominantly} $\pi^\pm$ and $\pi^0$, with a small admixture of $\omega^\pm$ and $\omega^0$ (at order $v^2/f^2$) as well as $\phi^\pm$ and $\phi^0$ (at order $v^{\prime 2}/v^2$). Likewise, the scalar fields absorbed by the $W_H$ and $B_H$ bosons are mostly $\omega$, but contain a small admixture of $\pi$. See Ref.~\cite{HMNP}.}. The masses and couplings of the $W$ and $Z$ bosons have the SM values in the limit $f\gg v$, but receive corrections of order $v^2/f^2$. These corrections are constrained by precision electroweak fits, which consequently put a lower bound on the scale $f$ (see section~\ref{pew}).
For generic values of $g_{1,2}, g_{1,2}^\prime$, the Coleman-Weinberg potential contains a coupling of the form $H^\dagger \phi H$; after the Higgs acquires a vev, this term generates a tadpole for $\phi$, which in turn induces a small but non-vanishing vev for the electrically neutral component of the triplet:
\beq
v^\prime\equiv \left< \phi_{22} \right> = -i \,\frac{v^2 f}{4M_\phi^2}\,
\left[ a \left(g_1^2+g_1^{\prime2}-g_2^2-g_2^{\prime2}\right)-a^\prime \lambda_1^2 \right].
\eeq{triplet_vev}
This additional contribution to EWSB violates custodial $SU(2)$ symmetry~\cite{cust}, and precision electroweak fits place strong constraints on its size.

\section{Alternative Realizations of the Little Higgs Mechanism}
\label{other}

While the Littlest Higgs model provides an explicit and economical  theory of EWSB based on the LH collective symmetry breaking mechanism, other interesting implementations of this mechanism have been proposed. It is useful to divide them into two classes~\cite{Simple0,Smoke}: ``product group'' models, in which the SM $SU(2)_L$ is embedded in a product gauge group, and ``simple group'' models, in which it is embedded in a larger simple group, e.g. $SU(3)$. The Littlest Higgs model belongs to the product group class; other examples from this class will be reviewed in subsection~\ref{product}. Particularly interesting are product group models with T parity~\cite{LHT0}, a discrete symmetry that alleviates experimental constraints on the model parameter space and has important phenomenological consequences. An example, the Littlest Higgs model with T parity, will be discussed in subsection~\ref{Tparity}. Finally, simple group models will be reviewed in subsection~\ref{simple}.

\subsection{Product Group Models}
\label{product}

Historically, the first realistic model utilizing the idea of collective symmetry breaking to describe EWSB was the ``big moose'' model~\cite{bigmoose}, constructed by Arkani-Hamed, Cohen and Georgi. A similar, but more economical ``minimal moose'' model was proposed in Ref.~\cite{minmoose}. The moose models embed the electroweak sector of the SM into a theory with a product {\it global} symmetry $G^N$, broken by a set of condensates, each of which transforms as a bifundamental representation under $G_i\times G_j$ for some pair of $i, j$. Such a structure can be described in terms of a   
``theory space''~\cite{decon,decon1} diagram. Such diagrams, which are a special case of the so-called ``moose diagrams''~\cite{moose}, consist of ``sites'' representing each symmetry factor $G_i$, and ``links'', connecting sites $i$ and $j$, representing bifundamental fields transforming under $G_i\times G_j$. A subset of the global symmetry is then gauged, including a factor of $SU(2)\times U(1)$, or a larger group containing $SU(2)\times U(1)$ as a subgroup, for each site. The bifundamental condensates break the gauge symmetry down to the SM $SU(2)\times U(1)$ at the TeV scale, with most, but not all, of the NGBs being absorbed by the heavy gauge bosons. Under appropriate conditions, the masses of some of the remaining NGBs are protected by the collective symmetry breaking mechanism. A general analysis~\cite{genmoose} shows that the protected NGBs are associated with topological properties of the theory space describing the model: each protected NGB corresponds to an element of the fundamental group of the theory space. Models in which the protected NGBs have the appropriate quantum numbers to be identified with the SM Higgs can be used to describe EWSB. Some generic phenomenological properties of such models have been discussed in~\cite{moosepheno}. 

\begin{figure}[t]
\begin{center}
\includegraphics[width=7cm]{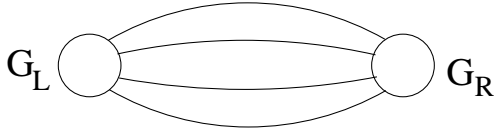}
\vskip2mm
\caption{The theory space diagram for the minimal moose LH model.}
\label{fig:minmoose}
\end{center}
\end{figure}

For example, the minimal moose model~\cite{minmoose} is described by the diagram in Fig.~\ref{fig:minmoose}. The gauge groups are $G_L=SU(2)\times U(1)$ at the left site, and $G_R=SU(3)$ at the right site. The SM fermions are charged under $G_L$ with their usual SM quantum numbers, and are singlets under $G_R$. Bifundamental condensates break the gauge symmetry down to the electroweak $SU(2)\times U(1)$. The dynamics of the theory is described by a gauged nlsm. The 
Lagrangian is constructed out
of gauge-invariant operators involving the four bifundamental link fields $X_j = \exp(2ix_j/f)$, where $x_j$ are the NGB fields, and covariant derivatives. (Note that $X_j^\dagger X_j=1$.) In addition to the usual kinetic terms for the 
$X_j$ fields, the symmetries of the model allow the so-called ``plaquette'' 
interactions, which contain no derivatives:
\beq
{\cal L}_{\chi} \,\sim\, f^4 \left( \Tr\left[ A_1 X_1 X_2^\dagger X_3 
X_4^\dagger\right] + {\rm ~cyclic}\right) \,+\, {\rm~h.c.}
\eeq{plaquette}
Here $A_i=\kappa_i+\eps_iT^8$, with $T^8$ denoting a diagonal $SU(3)$ 
generator and $\kappa_i$, $\eps_i$ dimensionless paremeters. These terms are 
required to obtain a stable, 
electroweak symmetry breaking minimum of the scalar potential in this model. 
The collective symmetry breaking mechanism can also be implemented in the top 
sector; the construction is analogous to the L$^2$H case discussed above.

The ``antisymmetric condensate'' model proposed in Ref.~\cite{antisym} is quite similar to the L$^2$H: an $SU(6)$ global symmetry is broken to an $Sp(6)$ subgroup by a condensate in the {\it antisymmetric} representation, proportional to
\beq
\Sigma_0^{\rm antisym} \,=\, \left(\begin{array}{cc}
0& -I\\
I& 0\\
\end{array}\right),
\eeq{cond}
where $I$ denotes a $3\times 3$ unit matrix. Again, an $[SU(2)\times U(1)]^2$ subgroup of the $SU(6)$ symmetry is gauged; this is broken down to the SM gauge group by the condensate~\leqn{cond}. After the breaking, 8 out of the original 12 Goldstone degrees of freedom remain uneaten; these transform as ${\bf 2}_{1/2}\oplus {\bf 2}_{-1/2}\oplus {\bf 1}_0$ under the SM $SU(2)_L\times U(1)_Y$. The masses of both doublets are protected by the collective symmetry breaking mechanism; as in the MSSM, there are two Higgs doublets at the weak scale. The Higgs potential, however, is distinctly different from that of the MSSM; in particular, the quartic terms have a different structure. This is reflected in the spectrum of the Higgs states: for example, in contrast to the MSSM this model predicts that the CP-odd scalar $A^0$ is heavier than the charged scalars $H^\pm$. The spectrum of the TeV-scale states is similar to the Littlest Higgs; however, the $B_H$ gauge boson tends to be somewhat heavier for the same value of $f$ due to a different normalization of the $U(1)$ generators in this model.

To avoid introducing large corrections to precision electroweak observables such as the $\rho$ parameter, the new physics at the TeV scale should respect (at least approximately) the custodial $SU(2)$ symmetry~\cite{cust}. The original L$^2$H model does not possess such a symmetry, and is severely constrained by precision electroweak fits. (This will be discussed in detail in section~\ref{pew}.) LH theories containing an approximate custodial $SU(2)$ have been constructed in Refs.~\cite{custodial,custodial_lh}. Unfortunately, these models are rather complicated: the model of~\cite{custodial} is a theory space model with four links, possessing an $[SO(5)]^8$ global symmetry and $SO(5)\times SU(2)\times U(1)$ gauge symmetry; the model of~\cite{custodial_lh} is based on an $SO(9)/[SO(4)\times SO(5)]$ nlsm. A different approach to improving the consistency with precision electroweak constraints, based on the introduction of a new discrete parity, will be discussed below.

\subsection{Littlest Higgs with T Parity}
\label{Tparity}

Early implementations of the Little Higgs mechanism turned out to be significantly constrained by precision electroweak observables (see section~\ref{pew}). To alleviate these constraints, Cheng and Low~\cite{LHT0,LHT} proposed to enlarge the symmetry structure of the models by introducing an additional discrete symmetry, dubbed ``T parity'' in analogy to R parity in the MSSM. T parity can be implemented in any LH model based on a product gauge group, including the Littlest Higgs~\cite{LHT1}. This parity explicitly forbids any tree-level contribution from heavy gauge bosons to observables involving only standard model particles as external states. In the case of the Littlest Higgs model, it also forbids the interactions that induced the triplet vev. As a result, in T parity symmetric Littlest Higgs
model, corrections to precision electroweak observables are
generated exclusively at loop level. This implies that the
constraints are generically weaker than in the tree-level case. In addition, as with R parity in the MSSM, the discrete symmetry ensures that the lightest T-odd particle (LTP) is stable. Typically, the LTP is the heavy partner of the hypercharge gauge boson (or, more colloquially, the ``heavy photon''), which provides a potential weak-scale dark matter candidate~\cite{JP}. In this subsection, we will review the Littlest Higgs model with T-parity (L$^2$HT).

The gauge sector of the model can be simply obtained from the original L$^2$H model reviewed in section~\ref{littlest}. In this sector,
T parity acts as an automorphism which exchanges the $\left[SU(2) \times U(1) \right]_1$ and $\left[SU(2) \times U(1) \right]_2$ gauge factors.  The Lagrangian in Eq.~\leqn{kinL} is invariant under this transformation provided that $g_1=g_2$ and $g^\prime_1=g^\prime_2$. In this case, the
gauge boson mass eigenstates (before EWSB) have the simple form, 
\beq
W_\pm = \frac{W_1 \pm W_2}{\sqrt{2}},~~~B_\pm = \frac{B_1 \pm B_2}{\sqrt{2}},
\eeq{WHBH}
where $W_+$ and $B_+$ are the standard model gauge bosons and are T-even, whereas $W_-$ and $B_-$ are the additional, heavy, T-odd states. (As already mentioned, $B_-$ is typically the 
lightest T-odd state, and plays the role of dark matter.) After  
EWSB, the T-even neutral states $W_+^3$ and $B_+$ mix to produce the SM $Z$ and the photon. Since they do not mix with the heavy T-odd states, the Weinberg angle is given by the SM relation, $\tan\theta_w=g^\prime/g$, where $g=g_{1,2}/\sqrt{2}$ and $g^\prime=g_{1,2}^\prime/\sqrt{2}$ are the SM gauge couplings, and $\rho=1$ at tree level. As will be shown below, all the SM fermions are also T-even, implying that the $W_-$ and $B_-$ states generate
no corrections to precision electroweak observables at all at tree level.

The transformation properties of the gauge fields under T parity and the
structure of the Lagrangian~\leqn{kinL} imply that T parity acts on 
the pion matrix as follows: 
\beq
T: \Pi \rightarrow - \Omega \Pi \Omega\,,
\eeq{Tpi}
where $\Omega = \mathrm{diag}(1,1,-1,1,1)$.  This transformation
law ensures that the complex $SU(2)_L$ triplet $\phi$ is odd under
T parity, while the Higgs doublet $H$ is even. The trilinear coupling of the 
form $H^\dagger \phi H$ is therefore forbidden, and no triplet vev is 
generated. Eliminating this source of tree-level custodial $SU(2)$ violation 
further relaxes the precision electroweak constraints on the model.

In the original L$^2$H model, the fermion sector of the standard model remained unchanged with the exception of the third generation of quarks, where the top Yukawa coupling had to be modified to avoid the large quadratically divergent contribution to the Higgs mass from top loops. In the model with T parity, however, the SM fermion doublet spectrum needs to be doubled to avoid
compositeness constraints~\cite{LHT}. For each SM lepton/quark doublet, two fermion doublets $\psi_1\in (\mathbf{2,1})$
and $\psi_2\in (\mathbf{1,2})$ are introduced. (The quantum numbers refer to 
representations under the $SU(2)_1\times SU(2)_2$ gauge symmetry.) These can be embedded in incomplete representations $\Psi_1,\Psi_2$ of the global $SU(5)$ symmetry. An additional set of fermions forming an $SO(5)$ multiplet $\Psi^c$, transforming nonlinearly under the full $SU(5)$, is introduced to give mass to the extra fermions; the field content can be expressed as follows:
\begin{equation}
\begin{array}{ccc}
\Psi_1=\left(\begin{array}{c} \psi_1 \\ 0 \\ 0 \end{array}\right)\,,
& \Psi_2=\left(\begin{array}{c} 0 \\ 0 \\ \psi_2
\end{array}\right) \,,&
\Psi^c=\left(\begin{array}{c} \psi^c \\ \chi^c \\ \tilde{\psi}^c
\end{array}\right).
\end{array}
\end{equation}
These fields transform under the $SU(5)$ as follows:
\beq
\Psi_1 \rightarrow V^* \Psi_1\,, \hspace{.2in} \Psi_2 \rightarrow V
\Psi_2\,, \hspace{.2in}\Psi^c \rightarrow U\Psi^c,
\eeq{su5}
where $U$ is the nonlinear transformation matrix defined in 
Refs.~\cite{LHT,LHT1,JP}. The action of T parity on the multiplets takes 
\beq
\Psi_1\leftrightarrow -\Sigma_0 \Psi_2, ~~~\Psi^c \to -\Psi^c.
\eeq{tpar}
These assignments allow a term in the Lagrangian of the form
\begin{equation}\label{heavyyuk}
\kappa f (\bar{\Psi}_2 \xi \Psi^c+\bar{\Psi}_1 \Sigma_0 \Omega
\xi^\dagger \Omega \Psi^c),
\end{equation}
where $\xi=\exp(i\Pi/f)$. This term gives a Dirac mass $M_-=\sqrt{2}\kappa f$ to the T-odd linear combination of $\psi_1$ and $\psi_2$, $\psi_-=(\psi_1+\psi_2)/\sqrt{2}$, together with $\tilde{\psi}^c$; the T-even linear combination, $\psi_+=(\psi_1-\psi_2)/\sqrt{2}$, remains massless and is identified with the standard model lepton or quark doublet. To give Dirac masses to the remaining T-odd states
$\chi^c$ and $\psi^c$, additional fermions with opposite gauge
quantum numbers can be introduced. For details, see Refs.~\cite{LHT,LHT1,JP}.

To complete the discussion of the fermion sector, we introduce the usual SM set of $SU(2)_L$-singlet right-handed leptons and quarks, which are T-even and can participate in the SM Yukawa interactions with the left-handed
$\psi_+$. The Yukawa interactions induce a one-loop quadratic divergence in the Higgs mass; however, the effect is numerically small except for the third generation of quarks. The Yukawa couplings of the third generation must be modified to incorporate the collective
symmetry breaking pattern. This requires completing the $\Psi_1$ and
$\Psi_2$ multiplets for the third generation to
representations of the $SU(3)_1$ (``upper-left corner'') and $SU(3)_2$ 
(``lower-right corner'') subgroups of $SU(5)$.  These multiplets are
\begin{equation}
\begin{array}{ccc}
{\cal Q}_1=\left(\begin{array}{c} q_1 \\ U_{L1} \\ 0 \end{array}\right) \,,&
{\cal Q}_2=\left(\begin{array}{c} 0 \\ U_{L2} \\ q_2
\end{array}\right);
\end{array}
\end{equation}
they obey the same transformation laws under T parity and the $SU(5)$ symmetry as do $\Psi_1$ and $\Psi_2$, see Eqs.~\leqn{su5} and~\leqn{tpar}. The quark doublets are embedded such that
\begin{equation}
q_{i} = -\sigma_2 \left(\begin{array}{c} u_{Li} \\ b_{Li}
\end{array}\right),~~~i=1,2.
\end{equation}
In addition to the SM right-handed top quark field $u_{3R}$, which is assumed to be T-even, the model contains two $SU(2)_L$-singlet fermions $U_{R1}$ and $U_{R2}$ of electric charge 2/3, which transform under T parity as
\begin{equation}
U_{R1}\leftrightarrow -U_{R2}.
\end{equation}
The top Yukawa couplings arise from the Lagrangian of the form
\beqa
{\cal L}_t &=& \frac{1}{2\sqrt{2}}\lambda_1 f \epsilon_{ijk}
\epsilon_{xy} \big[ ({\cal Q}_1^\dagger)_i \Sigma_{jx} \Sigma_{ky}  -
({\cal Q}_2^\dagger \Sigma_0)_i \tilde{\Sigma}_{jx} \tilde{\Sigma}_{ky}
\big] u_{3R} \arline && \hspace{1in} + \lambda_2 f (U^\dagger_{L1}
U_{R1} + U^\dagger_{L2} U_{R2})+ {\rm h.c.}
\eeqa{topyuk}
where $\tSigma=\Sigma_0\Omega \Sigma^\dagger \Omega
\Sigma_0$ is the image of the $\Sigma$ field under T parity, see Eq.~\leqn{Tpi}, and the indices $i,j,k$ run from 1 to 3 whereas $x,y=4,5$. The T parity eigenstates are given by
\beq
q_{\pm} = \frac{1}{\sqrt{2}}(q_1 \mp q_2),~~~~
U_{L\pm} = \frac{1}{\sqrt{2}}(U_{L1} \mp U_{L2}),~~~~
U_{R\pm} = \frac{1}{\sqrt{2}}(U_{R1} \mp U_{R2}).
\eeq{rot1}
The T-odd states $U_{L-}$ and $U_{R-}$ combine to form a Dirac fermion $T_-$,
with mass 
\beq
M_{T_-}=\lambda_2 f. 
\eeq{Tminusmass}
The remaining T-odd state $q_-$ receives a TeV-scale Dirac mass from the interaction in Eq.~\leqn{heavyyuk}. The Lagrangian for the T-even states is identical to the model without T parity, Eq.~\leqn{top_yuk}.  The T-even mass eigenstates are $t_+$, which acquires a mass only after EWSB and is identified with the SM top, and $T_+$, whose mass is equal to $\sqrt{\lambda_1^2+\lambda_2^2}f$, see Eq.~\leqn{heavytopmass}. The composition of $t$ and $T$ in terms of the original T-even fields is given in Eq.~\leqn{topmix}; the exact formulas including corrections of order $v/f$ can be found in Ref.~\cite{HMNP}. The cancellation of quadratic divergences in the top sector only involves T-even states, and occurs in precisely the same way as in the model without T parity. 

An alternative, more economical way to implement T parity in the top sector has recently been proposed in Ref.~\cite{Tnew}. In this approach, an additional weak doublet is not required. Three new weak-singlet Weyl fermions, $T$, $U^c_1$, and $U^c_2$, are introduced. Under T parity, $U^c_1\leftrightarrow U^c_2$ and $T\to -T$. Forming
a royal triplet $\chi=(b_L, u_L, T)$, the T-invariant Yukawa Lagrangian can be written as
\beq
{\cal L}_t \sim \lambda f  \left( \chi_i V_i U^c_1 \,+\, T[\chi]_i T[V]_i U^c_2\right),  
\eeq{top_newT}
where $V_i = \eps_{ijk}\eps_{xy} \Sigma_{jx}\Sigma_{ky}$, $T[\chi]=(b_L, u_L, -T)$, and $T[V]$ is the image of $V$ under T parity computed using Eq.~\leqn{Tpi}. The T-even linear combination of $U^c_1$ and $U^c_2$ is identified with the SM right-handed top quark, whereas the T-odd combination, together with $T$, acquire a Dirac mass of order $f$. Note that there are {\it no} new T-even states at the TeV scale in this model: {\it all} new particles are T-odd. The cancellation of the quadratically divergent top loop correction to the Higgs mass works analogously to the simple group models described in the next section: the cancellation is between the diagrams (a) and (c) in Fig.~\ref{fig:toploop}. The diagram (b) is absent, since the $tTh$ vertex is forbidden by T parity.

\subsection{Little Higgs from a Simple Group}
\label{simple}

The first model of the ``simple group'' type was constructed by Kaplan and Schmaltz~\cite{Simple0} (see also~\cite{Simple1}). Consider a theory with an $[SU(3)\times U(1)]^2$ global symmetry, spontaneously broken down to its $[SU(2)\times U(1)]^2$ subgroup by {\it two} vacuum condensates: 
\beq
\left<\Phi_{ ({\bf 3}, {\bf 1})} \right> \,=\, \left(\begin{array}{c}
0 \\ 0\\ f_1\\ \end{array}\right),~~~~~
\left<\Phi_{ ({\bf 1}, {\bf 3})} \right> \,=\, \left(\begin{array}{c}
0 \\ 0\\ f_2\\ \end{array}\right),  
\eeq{simple_vevs}
where $f_1\sim f_2 \sim 1$ TeV, and the subscripts indicate the $SU(3)\times SU(3)$ transformation properties of each condensate. The spontaneous global symmetry breaking gives rise to 10 Nambu-Goldstone bosons, whose low-energy dynamics are described in terms of the fields
\beq
\Phi_1(x) \,=\, e^{i \Theta_1(x)/f_1} \,\left<\Phi_{ ({\bf 3}, {\bf 1})} \right>\,,~~~ \Phi_2(x) \,=\, e^{i \Theta_2(x)/f_2} \,\left<\Phi_{ ({\bf 1}, {\bf 3})} \right>\,,
\eeq{sigma_simple}
where $\Theta_1$ and $\Theta_2$ are the pion matrices. 
The diagonal subgroup of the $[SU(3)\times U(1)]^2$ is gauged. 
The nlsm kinetic term is given by
\beq
{\cal L}_{\rm kin} \,=\, \sum_{i=1}^2 \left|\left(\partial_\mu + ig A_\mu^a T^a - i \frac{g_X}{3} B_\mu^{(X)} \right) \Phi_i \right|^2,
\eeq{kin_simple}
where $T^a$ are the $SU(3)$ generators, $A_\mu^a$ and $B_\mu^{(X)}$ are the $SU(3)$ and $U(1)$ gauge fields, respectively, and the $SU(3)$ gauge coupling constant $g$ is identical to the SM $SU(2)_L$ coupling. 
The condensates in Eq.~\leqn{simple_vevs} break the gauge symmetry down to the SM electroweak group $SU(2)_L \times U(1)_Y$, where the hypercharge group $U(1)_Y$ is identified with the unbroken linear combination of the $U(1)$ and the diagonal generator $T^8$ of $SU(3)$. Matching the SM hypercharge coupling constant requires
\beq
g_X \,=\,\frac{gt_w}{\sqrt{1-t_w^2/3}}\,.
\eeq{gXchoice}
where $t_w=s_w/c_w$ is the tangent of the SM Weinberg angle. Five of the nine gauge bosons acquire masses at the TeV scale, absorbing five NGB fields. In terms of the pion matrices, the absorbed combination has the form
\beq
\Psi(x) = \sin\theta \,\Theta_1(x) + \cos\theta\,\Theta_2(x),
\eeq{absorbed} 
where $\tan\theta=f_1/f_2$. The orthogonal combination, $\Theta(x)=\cos\theta \,\Theta_1(x)-\sin\theta \,\Theta_2(x)$, remains physical. The surviving NGBs decompose into a complex $SU(2)_L$ doublet, identified with the SM Higgs field $H=(h^0,h^-)$, and a SM singlet $\eta$: 
\beq
\Theta \,=\,  \left(\begin{array}{ccc}
0 & 0 & h^0\\ 0 & 0 & h^-\\ h^{0\dagger} & h^+ & 0\\ \end{array}\right)
\,+\, \frac{\eta}{\sqrt{2}}\,\left(\begin{array}{ccc}
1 & 0 & 0\\ 0 & 1 & 0\\ 0 & 0 & 1\\ \end{array}\right) \,.
\eeq{theta_def}
The TeV-scale gauge bosons in turn decompose into a complex $SU(2)_L$ doublet $(X^\pm, Y_{1,2}^0)$ and an $SU(2)_L$ singlet $Z^\prime$. The masses of these particles are given by~\cite{Smoke}
\beq
M_X=M_Y=\frac{gf}{\sqrt{2}}\approx0.46f,~~~
M_{Z^\prime}=\frac{\sqrt{2}gf}{\sqrt{3-t_w^2}}\approx 0.56f,
\eeq{mZp} 
where $f=\sqrt{f_1^2+f_2^2}$.
Note that all gauge boson masses are uniquely determined once the symmetry breaking scale $f$ is fixed.

In the SM, one-loop quadratically divergent contributions to the Higgs mass parameter arise from the bow tie diagrams, see Fig.~\ref{fig:bowtie}. The simple group LH model contains additional diagrams of this kind, with the TeV-scale gauge bosons running in the loop. In the gauge basis, the relevant couplings have the following structure:
\beq
\frac{g^2}{4}\,H^\dagger H\, \left[ 2 W^+_\mu W^{-\mu} + A^3_\mu A^{3\mu} - X^+_\mu X^{-\mu} - \frac{1}{2}\left(Y_{1\mu}^0 Y_1^{0\mu} + Y_{2\mu}^0 Y_2^{0\mu}\right) - A^8_\mu A^{8\mu}\right].
\eeq{simple_h2W2}
The cancellation of the one-loop quadratic divergence is manifest. (This cancellation can also be easily seen in the mass eigenbasis: recall that $W^\pm$, $X^\pm$ and $Y_{1,2}^0$ are already mass eigenstates, while the neutral mass eigenstates $\gamma$, $Z$ and $Z^\prime$ are linear combinations of $A^3$, $A^8$, and $B^{(X)}$.)
As in the product group models, this cancellation can be traced to the collective symmetry breaking mechanism, which, however, works in a slightly different way. The mechanism is best understood directly in terms of the fields $\Phi_1$ and $\Phi_2$. From the structure of the Lagrangian~\leqn{kin_simple}, it is clear that each bow tie diagram involves either a pair of $\Phi_1$ fields, or a pair of $\Phi_2$'s. The quadratic terms that such a diagram can generate must be proportional to either $\Phi_1^\dagger\Phi_1$ or $\Phi_2^\dagger\Phi_2$; but both these operators are identically equal to one, and do not involve any Higgs fields. Thus, the bow tie diagrams cannot induce a quadratically divergent Higgs mass. 
As in the L$^2$H model, one-loop logarithmic divergences are uncanceled: for example, the operator $|\Phi_2^\dagger \Phi_1|^2$, whose expansion contains a Higgs mass term, will be generated with a logarithmically divergent coefficient. The logarithmically divergent contribution from the top sector is generically dominant, and has the appropriate sign to trigger EWSB; one-loop finite and two-loop terms are subdominant. Note that the mass of the additional singlet scalar $\eta$ is also protected from the one-loop quadratic divergence, and this particle is expected to be light.

As in the L$^2$H model, the collective symmetry breaking mechanism should also be implemented in the top sector.
This is achieved by extending the third-generation quark doublet $q_{3L}$ of the SM to a ``royal'' triplet of the gauged $SU(3)$ group, $\chi_L^T = (q_{3L},U_L)$, and introducing two additional weak-singlet quarks, $U_{R1}$ and $U_{R2}$. These states are coupled by a Lagrangian of the form
\beq
{\cal L}_{\rm top} \,=\, \lambda_1 U^\dagger _{R1} \Phi_1^\dagger \chi_L \,+\, \lambda_2 U^\dagger _{R2} \Phi_2^\dagger \chi_L \,+\, {\rm h.c.} 
\eeq{simple_yuk}
The collective symmetry breaking mechanism works in this sector in the following way: the Lagrangian~\leqn{simple_yuk} breaks the global $SU(3)^2$ symmetry only when both $\lambda_1$ and $\lambda_2$ are non-zero. If, for example, $\lambda_2=0$, the field $\chi_L$ can be assumed to transform as {\bf (3,1)}, in which case the full $SU(3)^2$ is preserved. It follows that both $\lambda_1$ and 
$\lambda_2$ must appear in each top sector diagram that contributes to the Higgs mass renormalization, a condition that cannot be met by the one-loop quadratically divergent diagrams. 

One linear combination of the states $U_{R1}$ and $U_{R2}$ couples to $U_L$ to acquire a TeV-scale Dirac mass, whereas the orthogonal linear combination is identified with the right-handed top quark of the SM, and participates in the standard Yukawa interaction. To illustrate this, consider a simple case of symmetric vevs and couplings, $f_1=f_2=f/\sqrt{2}$, $\lambda_1=\lambda_2\equiv\lambda$. Expanding the $\Phi_{1,2}$ fields to quadratic order in $H$, we obtain
\beq
{\cal L}_{\rm t} \,=\, \lambda q_{3L}^\dagger H t_R + \lambda f U_L^\dagger 
U_R \left( 1-\frac{1}{2}\frac{H^\dagger H}{f^2}\right)\,+\,{\rm h.c.},
\eeq{simple_top}
where $t_R = (U_{R1}-U_{R2})/\sqrt{2}$, $U_R=(U_{R1}+U_{R2})/\sqrt{2}$. In diagramatic language, the cancellation of one-loop divergences from the top sector involves the same diagrams as in the Littlest Higgs case, see Fig.~\ref{fig:toploop}, except that diagram (b) is missing. The disappearance of this diagram, however, is special to the enhanced symmetry point we're considering. For example, if $\lambda_1\not=\lambda_2$ (but $f_1=f_2$), all three diagrams contribute, and the cancellation requires that the couplings obey the following relation:
\beq
\frac{m_T}{f}\,=\,\sqrt{\lambda_t^2+\lambda_T^2}.
\eeq{simple_sumrule}
A similar relation exists in the most general case, $f_1\not= f_2$, but it involves an additional parameter, the ratio $f_1/f_2$, and its structure is more complicated. Note that these relations are {\it different} from their Littlest Higgs counterpart, Eq.~\leqn{yuk_rel}. Thus, testing such relations experimentally could potentially allow one not just to obtain convincing evidence for the LH mechanism, but also to discriminate between different LH models. (The measurements needed to provide such a test at the LHC are discussed in section~\ref{collide}.) 

While there is no need to cancel quadratic divergences in the Higgs mass from light (non-top) quarks and leptons, additional states must be introduced to make the model theoretically consistent. The left-handed $SU(2)$ doublets of the SM must be embedded into triplets of the $SU(3)$ gauge group, $Q_a=(u_{La}, d_{La}, U_{La})^T$ and $L_a=(\nu_{La}, e_{La}, N_{La})^T$ where $a=1\ldots 3$ is the generation index. For each generation, two extra $SU(2)$ singlets, $U_{Ra}$ and $N_{Ra}$ must also be introduced in order to generate TeV-scale Dirac masses for the extra components of the triplets. This simplest ``universal'' embedding of the extra fermions, proposed in~\cite{Simple0}, is not anomaly-free, and additional fermions must appear at the scale $\Lambda\sim 10$ TeV to cancel the anomalies. An alternative ``anomaly-free'' embedding, in which all gauge anomalies are canceled at scale $f$, was proposed in Ref.~\cite{Kong}. This embedding has the same particle content, but the first two generations of quarks transform as a $\bar{\bf 3}$, instead of a {\bf 3}, under the $SU(3)$. The couplings of the light fermions and their TeV-scale partners, in both the universal and the anomaly-free embeddings, have been worked out in detail in Ref.~\cite{Smoke}.

In analogy to the L$^2$H model, the Higgs potential induced by quantum effects in the $SU(3)$ simple group LH model can be computed using the Coleman-Weinberg approach~\cite{CW}. As in the L$^2$H, logarithmically divergent one-loop diagrams induce the leading contributions to the Higgs mass parameter. The negative top contribution typically dominates due to a large Yukawa coupling, triggering EWSB. However, in contrast to the L$^2$H, the quartic coupling for the Higgs is {\it not} induced by quadratically divergent diagrams, and the predicted quartic coupling is too small to satisfy the lower bound on the Higgs mass. The simplest solution to this difficulty~\cite{Simple0,Simple1} is to simply add the required order-one quartic coupling by hand. This breaks the global symmetry in a way inconsistent with the Little Higgs mechanism, and induces a one-loop quadratic divergence in the Higgs mass from Higgs loops; however, this divergence is numerically quite small, and no major fine-tuning is required~\cite{Simple1}. A more satisfying solution~\cite{Simple0} is to extend the model to an $[SU(4)/SU(3)]^4$ nlsm with an $SU(4)\times U(1)$ gauged subgroup. In this case, the model contains two Higgs doublets, $H_u$ and $H_d$, and a coupling of the form $|H_u^\dagger H_d|^2$ is generated with no violation of the Little Higgs mechanism. The main disadvantage of this model is its complexity, with many new bosons and fermions necessarily appearing at the TeV scale.

An alternative simple group LH model, based on an $SU(9)/SU(8)$ global symmetry breaking pattern, has been proposed in Ref.~\cite{Simple2}. The electroweak gauge group is extended to $SU(3)\times U(1)$ and embedded in the $SU(9)$. The model contains two Higgs doublets at the electroweak scale, and an additional set of TeV-scale states.

\section{Precision Electroweak Constraints}
\label{pew}

The first test that any model postulating new physics at the TeV scale must pass is consistency with present experimental data. Precision measurements of numerous observables in the electroweak sector, performed over the last two decades, are especially constraining in this regard. The need to satisfy these constraints played an important role in the evolution of LH models. While the originally proposed models turned out to be tightly constrained, more recent constructions, such as the Littlest Higgs model with T parity, satisfy the constraints in large regions of the parameter space. This section will review the current status of precision electroweak constraints on a variety of Little Higgs models.

\subsection{Littlest Higgs}

\begin{figure}[t]
\begin{center}
\includegraphics[width=9.5cm]{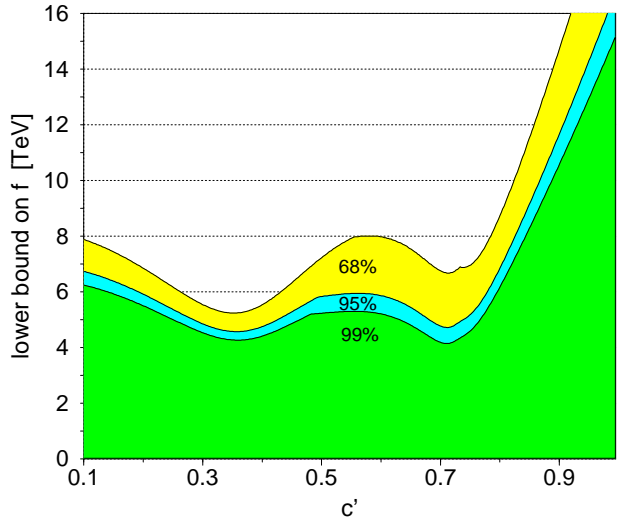}
\vskip2mm
\caption{Lower bound on the symmetry breaking scale $f$ in the $SU(5)/SO(5)$ Littlest Higgs model, as a function of the parameter $c^\prime\equiv \cos \psi^\prime$. The $SU(2)$ mixing angle $\psi$ is allowed to vary to give the least restrictive bound on $f$. (The theoretical constraint $\cos\psi\in[0.1, 0.995]$ is imposed to avoid the strong coupling regime.) From Ref.~\cite{CC1}.}
\label{fig:csaba}
\end{center}
\end{figure}

 Several studies of the corrections to precision electroweak observables (PEWO) induced in the Littlest Higgs model~\cite{CC1,HPR,Han,CD,Kilian,Marandella,Skiba} and its modifications~\cite{BPRS,PPP,CC2}) have been performed. The first analysis appeared in Ref.~\cite{CC1}. The model considered was the $SU(5)/SO(5)$ Littlest Higgs reviewed in section~\ref{littlest}. Light fermions were assumed to be charged under one of the $U(1)$ groups with the SM hypercharge, and neutral under the other $U(1)$. It was found that the leading corrections to PEWO are generated by the tree-level exchanges of heavy gauge bosons, explicit corrections due to non-linear sigma model dynamics, and the effects of the non-zero triplet scalar vev, Eq.~\leqn{triplet_vev}. In particular, weak isospin violating contributions were found to arise at tree level due to the absence of a custodial $SU(2)$ symmetry.
The bulk of these corrections results from heavy gauge boson exchanges, while a smaller contribution is due to the triplet scalar vev. A global fit to the experimental data was performed, and it was found that throughout the parameter space the symmetry breaking scale is bounded by $f > 4$ TeV at 95\% c.l. Even stronger bounds on $f$ were found for generic choices of the gauge couplings, see Fig.~\ref{fig:csaba}. The authors concluded that even in the best case scenario one would need fine tuning of less than a percent to get a Higgs mass as light as 200 GeV, largely destroying the original motivation for the L$^2$H model. Similar analyses, in the context of the same model, were presented in Refs.~\cite{HPR,Han,Kilian}. Their results generally agree with~\cite{CC1}. Refs.~\cite{Marandella,Skiba} update the analysis by including the constraints from LEP2 measurements of the  $e^+e^-\to f\bar{f}$ cross sections in the 189--207 GeV energy range, leading to an even tighter bound on $f$. 

In Ref.~\cite{CD}, one-loop corrections to the $\rho$ parameter were calculated, including the  logarithmically enhanced contributions from both fermion and scalar loops\footnote{The one-loop correction to the $Zb\bar{b}$ coupling was considered in Ref.~\cite{zbbar}. The correction to the muon anomalous magnetic moment was computed in Ref.~\cite{g-2} and found to be negligible.}. In certain regions of parameter space, the one-loop contributions were found to be comparable to tree level corrections, leading to modified constraints; in particular, partial cancellation of tree level and one-loop effects allowed for a somewhat lower bound on the scale $f$. 

The Littlest Higgs model contains an additional freedom of choosing the hypercharge assignments of the light fermions under the two $U(1)$ factors.  Ref.~\cite{CC2} addressed the question of whether a judicious choice of hypercharges can improve the consistency of the model with the PEWO. It was found that the situation was indeed improved, with values of $f$ as low as 
2 TeV being consistent with data; however, significant improvement only occurred in small regions of the model parameter space (e.g. the 
$\psi -\psi^\prime$ plane), and only for particular values of the parameter $R$ (see section~\ref{fermions}). Ref.~\cite{CC2} also considered alternative embeddings of the two $U(1)$ generators; again, the bound on $f$ could be lowered to 1--2 TeV, but only in small regions of the parameter space. 

\begin{figure}[t]
\begin{center}
\includegraphics[width=10cm]{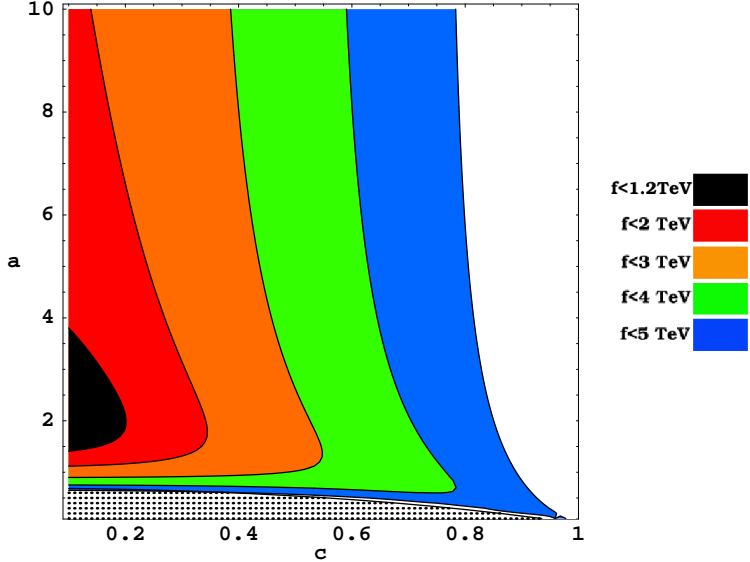}
\vskip2mm
\caption{Contour plot of the allowed values of $f$ in the variation of the $SU(5)/SO(5)$ Littlest Higgs model with an $SU(2)\times SU(2)\times U(1)_Y$ gauged subgroup, as a function of the parameters $c\equiv \cos \psi$ and $a$ (see Eq.~\leqn{CWg1}). The gray shaded region at the bottom is excluded by requiring a positive triplet mass. From Ref.~\cite{CC2}.}
\label{fig:oneU1}
\end{center}
\end{figure}

A more radical solution to the difficulties experienced by the original L$^2$H model was suggested by Arkani-Hamed and Wacker~\cite{private}. The large size of the precision electroweak corrections in this model is mostly due to the fact that the heavy hypercharge gauge boson, $B_H$, is actually typically not very heavy: according to Eq.~\leqn{WHmasses}, $M(B_H)\sim g^\prime f/\sqrt{5} \sim 0.16 f$ for generic parameters. Exchanges of such a light $B_H$ induce substantial corrections to $Z$ pole observables.
Moreover, a light $B_H$ is strongly disfavored by the null results of $Z^\prime$ searches at the Tevatron~\cite{HPR}. On the other hand, since the value of the hypercharge gauge coupling in the SM is rather small, $g^{\prime2}/g^2\sim0.3$, a model with a cutoff around 10 TeV does not require significant fine-tuning even if the one-loop quadratically divergent contributions to the Higgs mass from $B$ loops is {\it not} canceled. One can then consider a Littlest Higgs model with only one gauged $U(1)$ factor, i.e. the gauge group  
is $SU(2)_1\times SU(2)_2\times U(1)_Y$, in which the troublesome $B_H$ boson is absent. The precision electroweak corrections on this model have been analyzed in Refs.~\cite{CC2,PPP,BPRS}. It was found that, for certain values of model parameters, an $f$ as low as 1 TeV is allowed, see Fig.~\ref{fig:oneU1}. However, consistent fits with low $f$ are possible only if the mixing angle between the two $SU(2)$ groups, $\psi$, is close to $\pi/2$; that is, the relation $g_1\gg g_2$ is required. Unlike the original little hierarchy, this relation is technically natural; however, the model provides no explanation for its origin. Thus, the Littlest Higgs model with an $SU(2)\times SU(2)\times U(1)_Y$ gauged subgroup is consistent with experimental data, but somewhat unsatisfactory from a theoretical point of view.

Existing analyses of precision electroweak constraints on the Littlest Higgs (and other LH models) typically only include the calculable effects of weakly-coupled states at the scale $f$, ignoring the potential contributions from local operators generated at the cutoff scale $\Lambda$. This is justified as long as the expected hierarchy between the two scales, $f$ and $\Lambda\sim 4\pi f$, holds; however, an explicit analysis~\cite{He} of NGB scattering amplitudes in the L$^2$H model indicates a significantly smaller hierarchy, $\Lambda/f\sim 3-4$, due to the high multiplicity of NGBs. If this is the case, operators generated at $\Lambda$ may have substantial effects on precision electroweak fits; these effects remain to be investigated. 

\subsection{Alternative Little Higgs Models} 

In addition to the L$^2$H model, precision electroweak constraints on several alternative implementations of the Little Higgs mechanism have been studied in detail. They will be briefly reviewed in this subsection. Most studies presented their results in terms of the constraint on the global symmetry breaking scale $f$. We will follow this practice here; however, we note that a comparison of the lower bounds on $f$ obtained in different models does {\it not} provide a reliable measure of the relative amount of fine tuning they require, since the precise relation between $f$ and the Higgs mass induced by quantum loops is itself model-dependent. 

Precision electroweak constraints on the $SU(6)/Sp(6)$ antisymmetric condensate model were considered in Refs.~\cite{CC2,PEW66,Marandella,Skiba}. The constraints were found to be quite similar to those in the Littlest Higgs: in particular, in the ``minimal'' implementation of the model (with two gauged $U(1)$ factors and $R=0$ or 1), the scale $f$ has to be above 3.0 TeV. Again, modifying the way in which the gauged $U(1)$ generators are embedded, modifying the fermion $U(1)$ charges, or gauging a single $U(1)$, opens up the possibility of $f$ as low as 1 TeV, albeit only in rather small regions of the parameter space. Ref.~\cite{CC2} has also analyzed the constraints on the $[SU(4)/SU(3)]^4$ simple group model (see section~\ref{simple}), and obtained a lower bound of $f>4.2$ TeV. While this bound is tighter than what can be achieved in the product-group models, large $f$ in these models does not necessarily imply a large amount of fine tuning, since the heavy fermions which cancel the divergence in the top loop contribution to the Higgs mass parameter may appear at a scale substantially below $f$ for a carefully chosen set of parameters. Ref.~\cite{Simple1} studied the simpler $SU(3)$ simple group Little Higgs model, with an explicit Higgs quartic coupling introduced by hand, and found that this model has a region of parameter space in which EWSB is natural and the precision electroweak constraints are satisfied. The analysis of this model in Ref.~\cite{Marandella}, however, finds a much more stringent constraint, $f\gapproxeq 4.5$ TeV. (A similar constraint has been found in Ref.~\cite{Skiba}.) The constraints on the $SU(9)/SU(8)$ model were considered in Refs.~\cite{Simple2,Marandella}; in this model, the bound on $f$ is 3.3 TeV, but in large parts of the parameter space this is compatible with a heavy top mass (the scale at which the largest quadratically divergent contribution to the Higgs mass is canceled) below 2 TeV. Precision electroweak observables in the minimal moose model were analyzed in Ref.~\cite{PEWmin}. It was found that the original model of~\cite{minmoose} is very tightly constrained. However, a slightly modified version of the model, with the $SU(3)\times SU(2)\times U(1)$ gauge group replaced by $[SU(2)\times U(1)]^2$, is significantly less constrained, allowing for $f$ as low as 2 TeV. Finally, the constraints on the models with custodial symmetry~\cite{custodial,custodial_lh} were considered in Ref.~\cite{custPEW}. This analysis included the corrections to the muon anomalous magnetic moment as well as the precision electroweak observables. While no quantitative bounds have been presented, the analysis concluded that the introduction of custodial symmetry does lead to a substantial reduction in the amount of fine-tuning. 

\subsection{Littlest Higgs with T Parity}
\label{pewT}

\begin{figure}[t]
\begin{center}
\includegraphics[width=8cm]{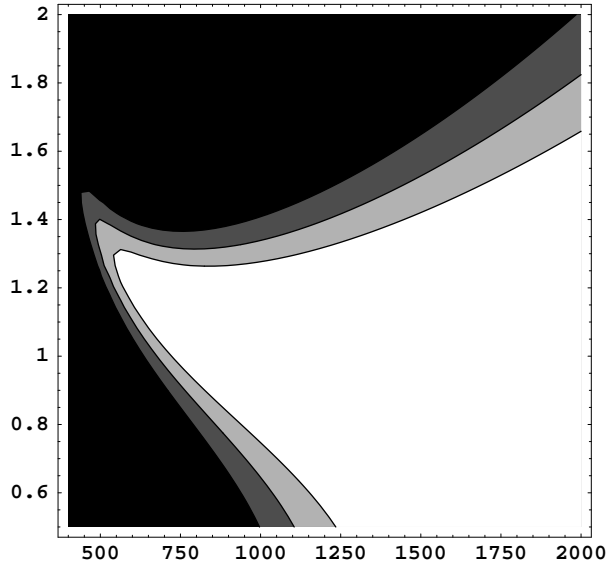}
\vskip2mm
\caption{Exclusion contours in terms of the parameter $R=\lambda_1/\lambda_2$ 
and the symmetry breaking scale $f$. The contribution of the T-odd fermions 
to the T parameter is neglected. From lightest to darkest, the contours 
correspond to the 95, 99, and 99.9 confidence level exclusion.
From Ref.~\cite{HMNP}.}
\label{fig:TparEW}
\end{center}
\end{figure}

Motivated by the difficulties in reconciling the original Littlest Higgs and other early proposals with precision electroweak observables, Cheng and Low proposed to extend the symmetry of the models to include a discrete T parity. An explicit example of how this can be achieved, the L$^2$H model with T parity, has been reviewed above (see section~\ref{Tparity}).  The parity explicitly forbids any tree-level contribution from heavy gauge bosons to observables involving only SM particles as external states. In the case of the L$^2$H model, it also forbids the interactions that induced the triplet vev. As a result, in this
model, corrections to precision electroweak observables are
generated exclusively at loop level. This implies that the
constraints are generically weaker than in the tree-level case. A quantitative analysis of the constraints was presented in Ref.~\cite{HMNP}. The oblique corrections~\cite{PT} from loops of T-odd particles, as well as the T-even $T_+$, were computed, along with the $Zb\bar{b}$ vertex correction from the top sector. It was found that large regions of parameter space are consistent with precision electroweak constraints, see Fig.~\ref{fig:TparEW}. In particular, consistent fits were obtained for $f$ as low as 500 GeV, indicating that the model does not require significant amounts of fine tuning to keep the Higgs light. Interestingly, the model also allows for consistent fits with a heavy Higgs boson, up to 800 GeV, since the large negative contribution to the T parameter induced by a heavy Higgs can be cancelled by a positive contribution from $T_+$ loops. In addition, an {\it upper} bound on the mass of the T-odd fermions was obtained from the non-observation of non-SM four-fermion interactions, such as $eedd$, in collider experiments. Assuming a flavor-universal T-odd fermion mass, it was found 
that
\beq
M_{\rm TeV} < 4.8 f^2_{\rm TeV}\,,
\eeq{upperT}
where $M_{\rm TeV}$ and $f_{\rm TeV}$ are values of the T-odd fermion mass and the symmetry breaking scale, respectively, in TeV. 
A more detailed analysis of the low-energy effects of the T-odd fermions, including the flavor-violating effects induced if their masses and couplings are not universal, remains to be performed.

\section{Collider Phenomenology}
\label{collide}

In order to cancel the one-loop quadratic divergences in the Higgs mass, all Little Higgs theories require new particles at the TeV scale. Independent of the specific model, the TeV-scale spectrum includes a vector-like quark, required to cancel the top loop divergence, and 
a set of new gauge bosons, canceling the $W/Z$ loop divergences. 
Moreover, the symmetries of the LH theory relate the couplings of these particles with the Higgs to the SM gauge and Yukawa couplings (see, for example, Eqs.~\leqn{higgs_coupl1} and~\leqn{yuk_rel}). These are the generic predictions of the collective symmetry breaking mechanism, although the detailed form of the coupling relations is somewhat model-dependent. The masses of the new particles are bounded from above by naturalness considerations. While the precision electroweak constraints make a discovery at the Tevatron rather unlikely (with the possible exception of models with T parity), at least some part of the spectrum should be observable at the LHC. Specific LH models often contain additional TeV-scale particles (scalars, fermions, and gauge bosons), which are not required by the LH mechanism itself but are needed to implement it in a theoretically consistent way. Observing these particles could provide important hints to which particular LH model is realized in nature. This section will review the prospects for discovery and study of the new particles predicted by LH models at future colliders. 

\subsection{Heavy Gauge Boson Phenomenology}

The study of the collider phenomenology in a Little Higgs model was initiated in Refs.~\cite{HPR,BPP} (see also~\cite{Han}), which considered the signatures of the extra gauge bosons in the context of the Littlest Higgs model. The model predicts four new gauge bosons, $W_H^\pm$, $W_H^3$ and $B_H$, with masses given in Eq.~\leqn{WHmasses}. At a hadron collider, these bosons are produced predominantly through their coupling to quarks. Assuming that the SM left-handed lepton and quark doublets $L$ and $Q$ transform as doublets under $SU(2)_1$ and singlets under $SU(2)_2$, their couplings to the heavy $SU(2)$ bosons are flavor-universal and have the form\footnote{Another simple and consistent $SU(2)$ charge assignment corresponds to interchanging $SU(2)_1\leftrightarrow SU(2)_2$; the results for this case can be obtained by exchanging 
$\psi\leftrightarrow \pi/2-\psi$ in Eq.~\leqn{fermion_c} and below.}
\beq
g \cot\psi\,W^a_{H\mu}\,\left( \bar{L} \gamma^\mu \frac{\sigma^a}{2}\,L\,+\,
\bar{Q} \gamma^\mu \frac{\sigma^a}{2}\,Q\right).
\eeq{fermion_c}
The coupling of the $B_H$ boson to the SM fermions depends on the fermions' $U(1)_1$ and $U(1)_2$ charges, and is quite model-dependent. In fact, eliminating the $B_H$ altogether may reduce the amount of fine tuning in this model when the precision electroweak constraints are taken into account, see section~\ref{pew}. Given this model uncertainty in the $U(1)$ sector, the analysis of Ref.~\cite{BPP} focused on the production and decay of the $SU(2)$ heavy gauge bosons.

\begin{figure}[t]
\begin{center}
\epsfig{file=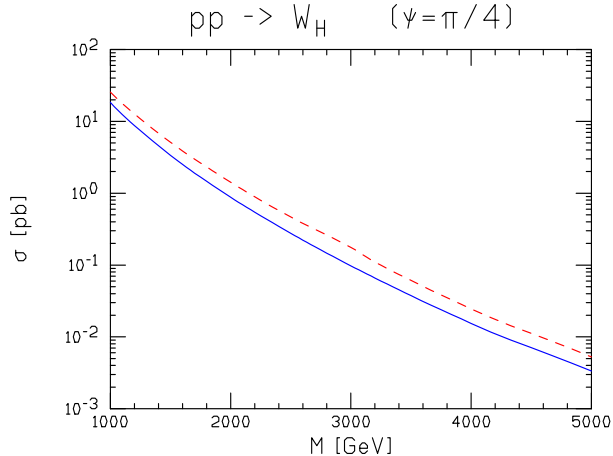,width=6cm,height=8cm,angle=90}
\vskip2mm
\caption{Production cross sections for the $W^3_H$ (solid) and $W_H^\pm$ (dashed) bosons at the LHC, for $\psi=\pi/4$. From Ref.~\cite{BPP}.}
\label{fig:w3prod}
\end{center}
\end{figure}

Precision electroweak constraints on the Littlest Higgs model imply that the $W_H$ bosons are out of the kinematic reach of the Tevatron collider. At the LHC, these bosons would be mostly 
produced through $q \bar{q}$ annihilation. (The sub-process $q g \to W_H q$ has a considerably smaller cross section and could be separately identified due to the presence of a high $p_T$ jet.) Fig.~\ref{fig:w3prod} shows the leading order production cross section of $W_H^3$ and $W^\pm_H$ as a function of their mass, for the case $\psi=\pi/4$. The general case may be obtained from Fig.~\ref{fig:w3prod} by simply scaling by $\cot^2\psi$. The decay channels of the $W_H^3$ boson include $\ell^+\ell^-$, $q\bar{q}$, $Zh$, $W^+W^-$, and, potentially, $B_H h$. Ignoring the model-dependent $B_Hh$ mode, the total width of $W_H^3$ is given by
\beq
\Gamma_{\rm tot} = \frac{g^2}{96\pi}\,\left(\cot^2 2\psi + 24 \cot^2\psi
\right)\,M, 
\eeq{total_w}
where $M\equiv M(W_H)$. The partial widths are
\beqa
\Gamma(W_H^3\to \ell^+\ell^-) = \frac{g^2 \cot^2\psi}{96\pi}\,M\,,~&~&~
\Gamma(W_H^3\to \bar{q}q) = \frac{g^2 \cot^2\psi}{32\pi}\,M, \CR
\Gamma(W_H^3\to Zh) = \frac{g^2 \cot^2 2\psi}{192\pi}\,M,
& &\Gamma(W_H^3\to W^+W^-) = \frac{g^2 \cot^2 2\psi}{192\pi}\,M.
\eeqa{widths}
where we neglect corrections of order $v/f$ (including the effects of non-zero top mass). Partial decay widths of the $W_H^{\pm}$ bosons are easily obtained from~\leqn{widths} using the isospin symmetry, which is accurate to leading order in $v/f$.  

The discovery reach of the LHC for the $W^3_H$ and $W^{\pm}_{H}$ gauge bosons is quite high. The cleanest mode is $W^3_H\to\ell^+\ell^-$, with $\ell=e$~or~$\mu$. The studies of the $Z^\prime$ discovery reach at the LHC performed by the ATLAS collaboration~\cite{atlas} indicate that these channels are virtually free of backgrounds. The discovery reach (corresponding to the observation of $10$~events) for 100~fb$^{-1}$ integrated luminosity is approximately $M(W_H)=5\cdot(\cot \psi)^{1/3}~$TeV. 
The same decay modes will provide a determination of the $W_H$ mass. However, discovering the $W_H$ triplet does not by itself provide a striking signature for the LH model, since one can imagine many alternative theories in which such a triplet is present.

\begin{figure}[t]
\begin{center}
\epsfig{file=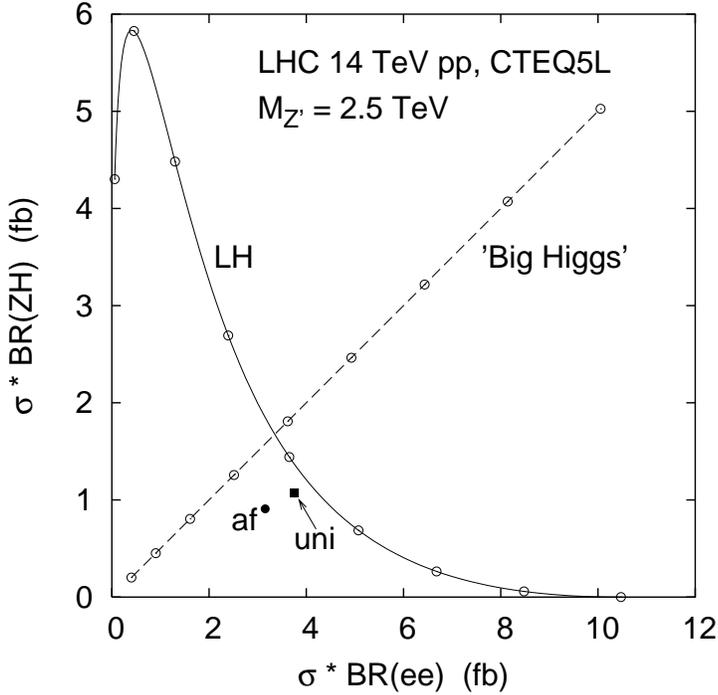,width=9.5cm,angle=-90}
\vskip2mm
\caption{Predictions of the $ee$ and $Zh$ event rates due to $W^3_H$ decay at the LHC, in three Little Higgs models (Littlest Higgs, LH; simple group Little Higgs with two alternative fermion embeddings, ``af'' or anomaly-free and ``uni'' or universal, see section~\ref{simple}) and a non-Little Higgs model with the same spectrum (Big Higgs). From Ref.~\cite{Smoke}.}
\label{fig:smoke1}
\end{center}
\end{figure}

\begin{figure}[t]
\begin{center}
\epsfig{file=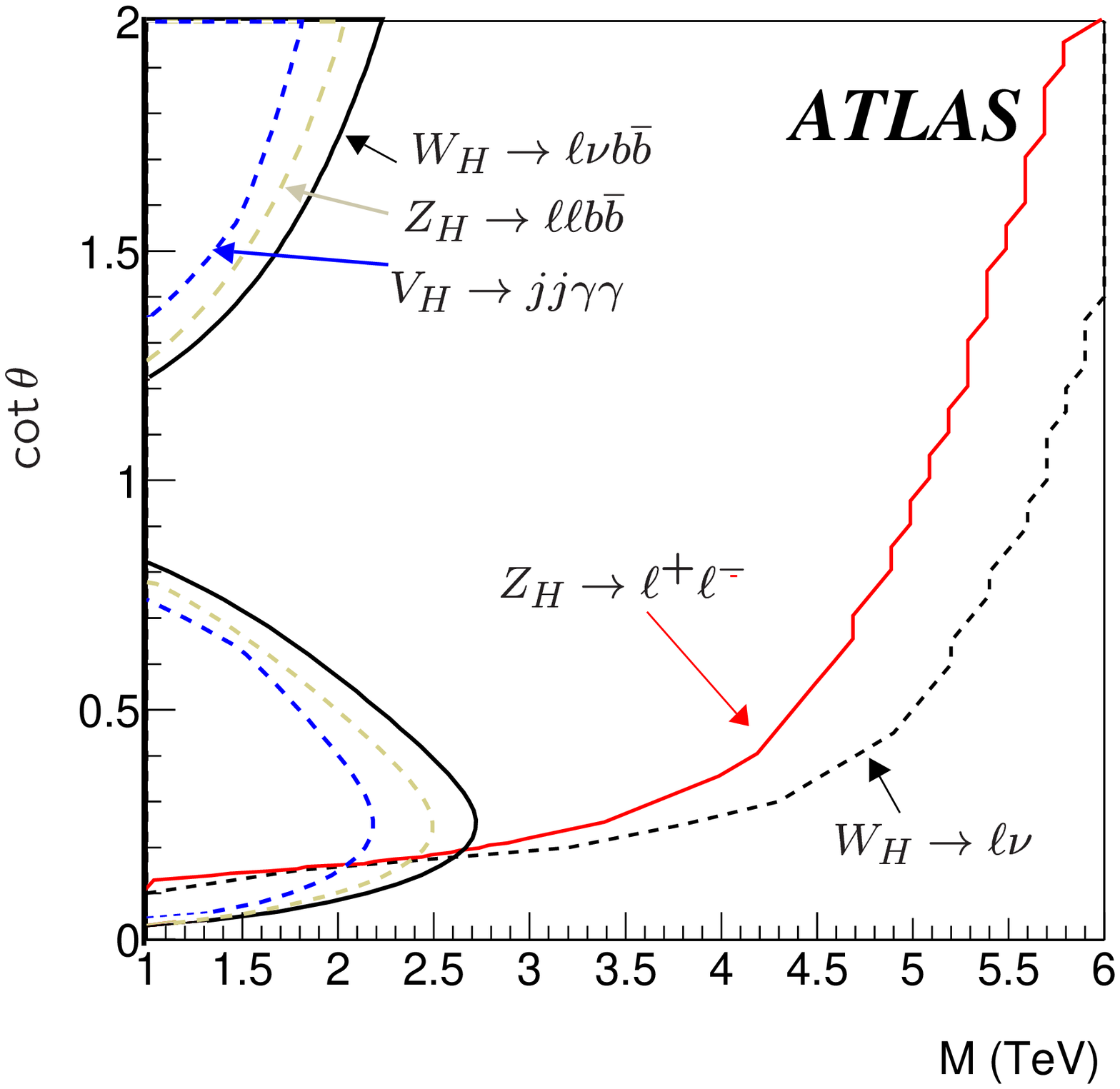,width=9.5cm,angle=0}
\caption{Accessible regions, in the $M(W_H)-\cot\psi$ plane, for $5\sigma$ discovery at the LHC with 300 fb$^{-1}$ integrated luminosity. 
From Ref.~\cite{ATLAS}.}
\label{fig:ATLAS_W}
\end{center}
\end{figure}

A clean way to verify the L$^2$H origin of the observed $W_H$ bosons experimentally has been proposed in Ref.~\cite{BPP}. The key observation is that the factor of $\cot 2\psi$ in the $W_H Z h$ coupling is a unique feature of the L$^2$H model. This coupling originates from the $W_H W_L H^\dagger H$ term in Eq.~\leqn{higgs_coupl1}, whose coefficient is determined by the special structure of Eq.~\leqn{higgs_coupl}, which in turn is a direct consequence of the collective symmetry breaking pattern. Therefore, an independent determination of the mixing angle $\psi$ and the $W^3_H \to Zh$ (and/or $W_H^\pm \to W^\pm h$) partial widths would provide a robust test of the L$^2$H model. Assuming that the production cross section can be accurately predicted, such a determination is provided by simply counting the number of events in the $\ell^+\ell^-$ and $Zh$ channels with the invariant mass equal to $M(W_H)$. Ref.~\cite{BPP} considered a sample toy model, the ``Big Higgs'', possessing the same spectrum as the L$^2$H model but without the collective symmetry breaking mechanism. It was shown that the measurements outlined above allow one to distinguish between these models. A recent analysis of Ref.~\cite{Smoke} has confirmed this conclusion, see Fig.~\ref{fig:smoke1}. Note, however, that this strategy relies on an accurate prediction of the $W_H$ production cross section, which in turn requires precise knowledge of quark and antiquark distribution functions at $x\sim 0.1-0.2$. 

A more detailed analysis of the $W_H$ signatures in the $W^3_H\to\ell^+\ell^-, Zh$, and $W_H^\pm\to \ell^\pm\nu, W^\pm h$ channels, including the backgrounds and detector effects, 
has been performed by the ATLAS collaboration~\cite{ATLAS}. In the $Zh$ channel, the signatures studied included $\ell^+\ell^- b\bar{b}$ and $jj\gamma\gamma$; in the $W^\pm h$ channel, $\ell^\pm\nu b\bar{b}$ and $jj\gamma\gamma$ signatures were considered. (The study assumed a Higgs mass of 120 GeV.) The results are summarized in Figure~\ref{fig:ATLAS_W}, where the discovery reaches for various final states are shown. All of the natural parameter range is covered by the $\ell^+\ell^-$, $\ell^\pm \nu$ signatures; for lower values of $M(W_H)$, several decay channels are accessible, allowing the model to be tested.

The International Linear Collider (ILC), a proposed $e^+e^-$ collider with center-of-mass energy $\sqrt{s}=500$ GeV at the initial stage and up to 1 TeV after upgrades, is not likely to have sufficient energy to produce the $W_H$ bosons of the L$^2$H model directly. (It is possible that the $B_H$ boson, if present, could be kinematically accessible at the 1 TeV ILC and produced via $e^+e^-\to B_H$~\cite{ILC_BHdirect} or $e\gamma \to eB_H$~\cite{egamma}; also, some of the allowed parameter space in models with T parity could be accessible.) However, precise measurements of scattering cross sections at the ILC will allow us to detect and study these bosons via their indirect effects even if $M(W_H)\gg\sqrt{s}$. In fact, it was found in Ref.~\cite{ILC1} that the search reach for the $W_H^3$ and $B_H$ bosons in the process $e^+e^-\to f\bar{f}$ at a 500 GeV ILC covers essentially the entire parameter space of the Littlest Higgs model consistent with naturalness\footnote{The effects of the $B_H$ boson on the $e^+e^-\to \mu^+\mu^-$ process have been considered in Ref.~\cite{Park}, while the process $e^+e^-\to t\bar{t}$ was analyzed in Ref.~\cite{ILCttbar}.}. Moreover, studying this channel at the ILC, combined with the measurement of the $W_H$ (and possibly $B_H$) masses at the LHC, allows a determination of the fundamental parameters of the model ($f$, $\psi$, and $\psi^\prime$) to the precision of a few per 
cent. Unlike the LHC parameter determination, this method is not limited by the parton distribution function uncertainties.
In a significant region of the parameter space, the effect of the $B_H$ boson on the process $e^+e^-\to Zh$ can also be observed~\cite{ILC_ZH,ILC1}. This process may allow one to extract the $W_H^3Zh$ coupling, which would provide an unambiguous test of the L$^2$H model as explained above. 

\begin{figure}[t]
\begin{center}
\includegraphics[width=7cm]{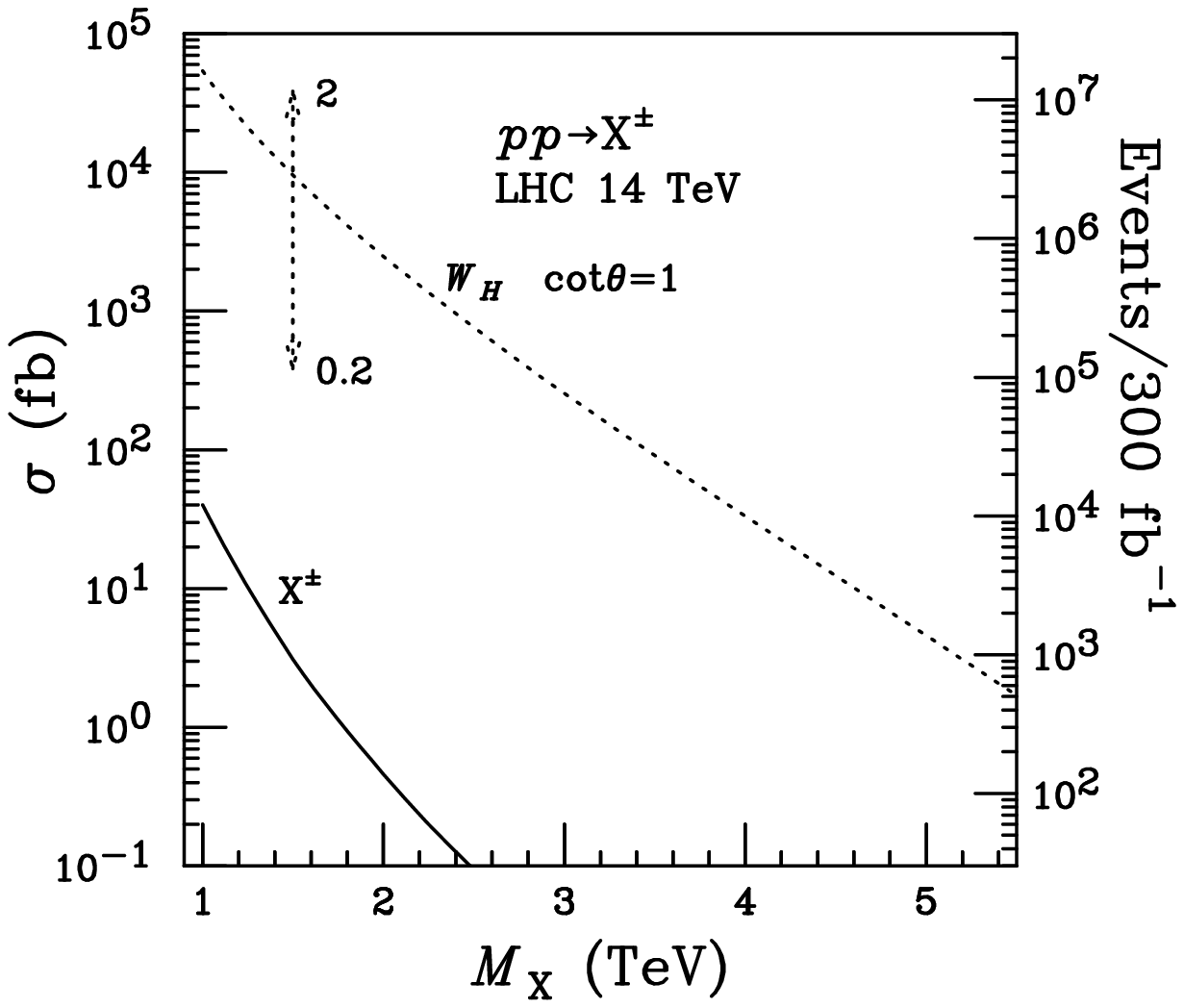}
\includegraphics[width=7cm]{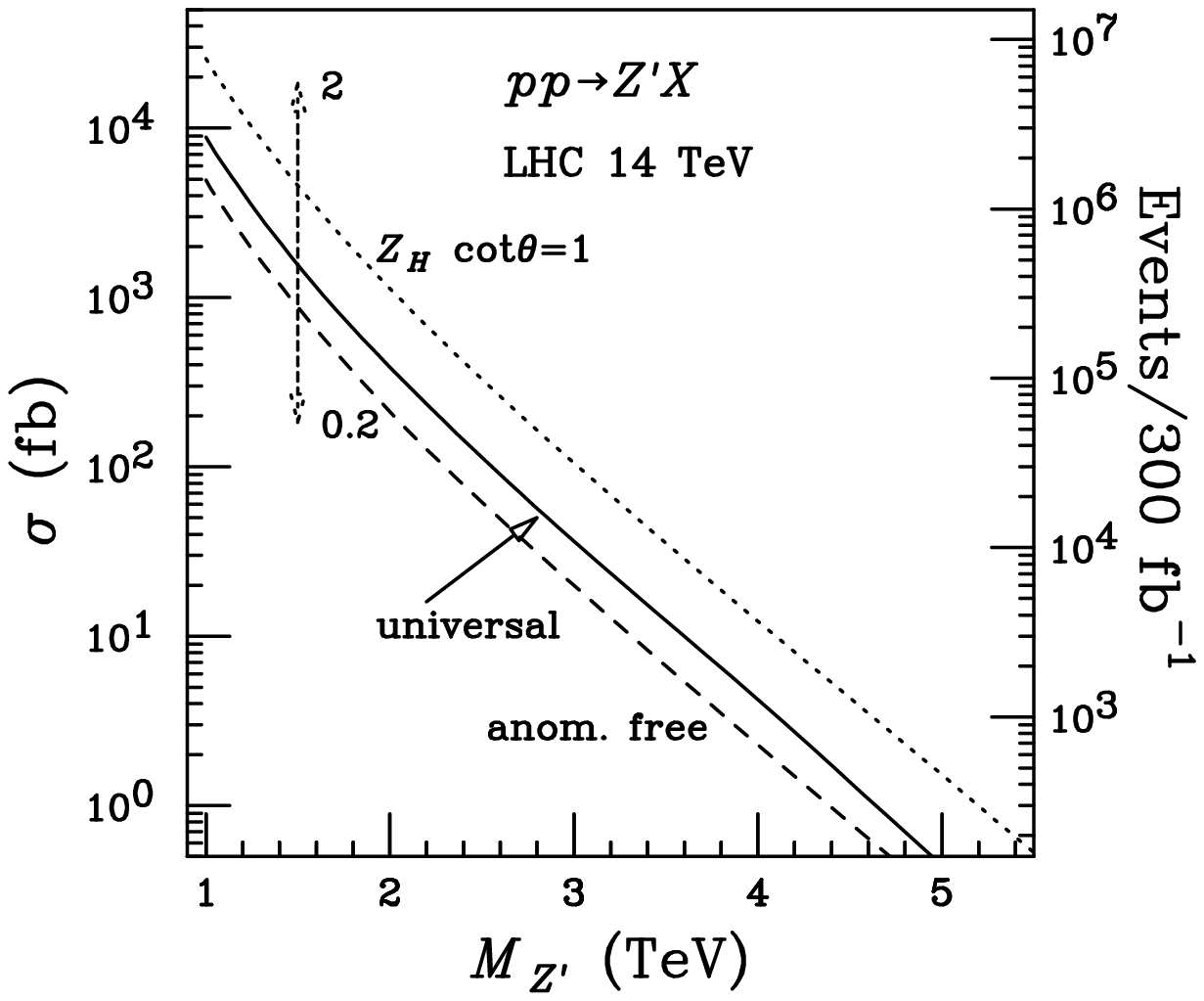}
\vskip2mm
\caption{Production cross sections for the TeV-scale neutral (left panel) and charged (right panel) gauge bosons of Little Higgs models at the LHC. Solid lines correspond to the $SU(3)$ simple group model with universal fermion embedding, and the dashed line on the left panel to the same model with an alternative (``anomaly-free'') fermion embedding. Dotted lines represent the Littlest Higgs model with $\psi=\pi/4$, and dotted arrows show the variation of the Littlest Higgs $W_H^3$ production cross section when $\cot\psi$ is varied between 0.2 and 2. From Ref.~\cite{Smoke}.}
\label{fig:smoke2}
\end{center}
\end{figure}

The only study so far of collider phenomenology in an LH model of the simple group type has appeared in Ref.~\cite{Smoke}, where the signatures of the $SU(3)$ model (see section~\ref{simple}) were analyzed and contrasted with those expected in the Littlest Higgs. The gauge sector of the $SU(3)$ model contains two new complex gauge bosons, $X^\pm$ and $Y$, as well as a neutral $Z^\prime$; the masses of these states are given in Eq.~\leqn{mZp}. Interestingly, this sector contains no dimensionless parameters apart from the known SM gauge couplings, so that the properties of the heavy gauge bosons are uniquely specified up to an uncertainty in the scale $f$ and a discrete ambiguity in the fermion embedding. The $X$ and $Y$ bosons do not couple to the SM fermions in the limit $v/f\to 0$; the production cross sections for these states are suppressed by $v^2/f^2$, and they will most likely not be observable at the LHC. This is illustrated in the right panel of Fig.~\ref{fig:smoke2}, where the production cross section for the $W_H^\pm$ of the L$^2$H model is also shown for comparison. On the other hand, the coupling of the $Z^\prime$ boson to the SM fermions is unsuppressed, and the production cross section is substantial, as shown on the left panel of Fig.~\ref{fig:smoke2}. The $SU(3)$ simple group model also makes unambiguous predictions of the $Z^\prime$ partial decay widths. As illustrated in Fig.~\ref{fig:smoke1}, measuring the $\ell^+\ell^-$ and $Zh$ event rates would enable one to distinguish this model both from non-LH $Z^\prime$ models and from its Littlest Higgs cousin, even taking into account the presence of the tunable parameter (the mixing angle $\psi$) in the L$^2$H model.

Collider phenomenology of the Littlest Higgs model with T parity was considered in Ref.~\cite{JP}. While the gauge boson spectrum of this model is the same
as in the original Littlest Higgs (with the restriction $\psi=\psi^\prime=\pi/4$), the phenomenology is drastically different due to the fact that the TeV-scale gauge bosons are T-odd.
In this sense, the expected signatures resemble those of a SUSY model with conserved R parity: for example, since all SM particles are T-even, the heavy gauge bosons must be pair-produced. The $B_H$ gauge boson, whose presence is obligatory in this model, is quite light, $M(B_H)=g^\prime f/\sqrt{5}\approx 0.16 f$, and is typically the lightest T-odd particle (LTP). If T parity is exact, the LTP is stable, and events with $W_H$ or $B_H$ pair-production will be characterized by large missing transverse momentum carried away by the two LTPs.

\subsection{Heavy Top Phenomenology}

\begin{figure}[t]
\begin{center}
\includegraphics[width=9.5cm]{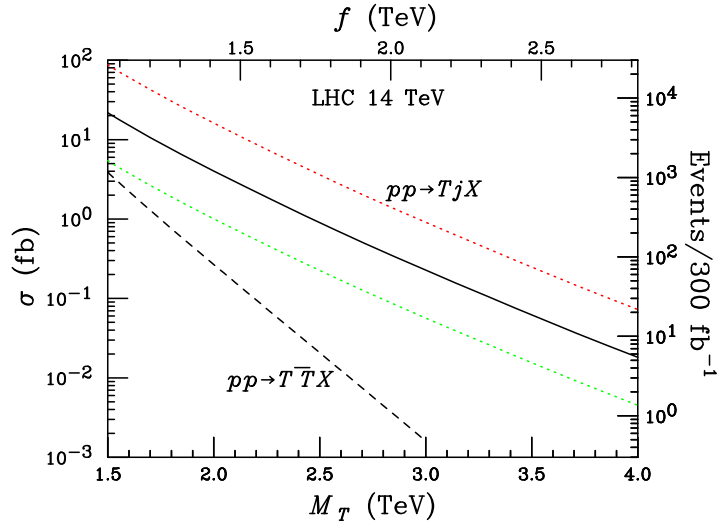}
\vskip2mm
\caption{Production cross section of the $T$ quark at the LHC. The solid and lower and upper dotted lines correspond to the single $T$ production for $\lambda_1/\lambda_2=1$, $1/2$, and $2$, respectively. The dashed line corresponds to the $T\bar{T}$ pair production.  From Ref.~\cite{Han}.}
\label{fig:Tprod}
\end{center}
\end{figure}

A new vector-like colored fermion at the TeV scale is a generic prediction of LH models, since its presence is required to cancel the one-loop divergence from the top loops in Fig.~\ref{fig:smtoploop}. Naturalness indicates that the mass of this fermion, which is usually referred to as the ``$T$ quark'' or the ``heavy top'', should be below about 2 TeV. The collider phenomenology of the $T$ quark in the context of the Littlest Higgs model has been studied in Refs.~\cite{Han,PPP}. At the LHC, the $T$ quark can be pair-produced via processes such as $gg\to T\bar{T}$ and $q\bar{q}\to T\bar{T}$, or singly produced via the $W$ exchange process $bq\to Tq^\prime$. The cross sections of these processes are shown in Fig.~\ref{fig:Tprod}. The pair production, due to strong interactions, dominates for low $T$ masses; however for $M_T$ above about 1 TeV, energy is at a premium and single production becomes dominant. The decay pattern of the $T$ can be understood from Eq.~\leqn{tophiggs} by applying the Goldstone boson equivalence theorem: the $T$ couples symmetrically to the four components of the Higgs doublet, leading, in the $M_T\gg v$ limit, to the following simple prediction:
\beq
\Gamma (T \to t h) = \Gamma(T\to t Z^0) = \half \Gamma(T\to b W^+)\,=\, \frac{\lambda_T^2 M_T}{64\pi}\,.
\eeq{Tdecays}
Additional decay modes, involving the TeV-scale gauge bosons of the Littlest Higgs model, may be kinematically allowed and contribute to the total $T$ width: for example, if the $B_H$ boson is present and light, the decay $T\to tB_H$ may be possible. 

\begin{figure}[t]
\begin{center}
\includegraphics[width=9.5cm]{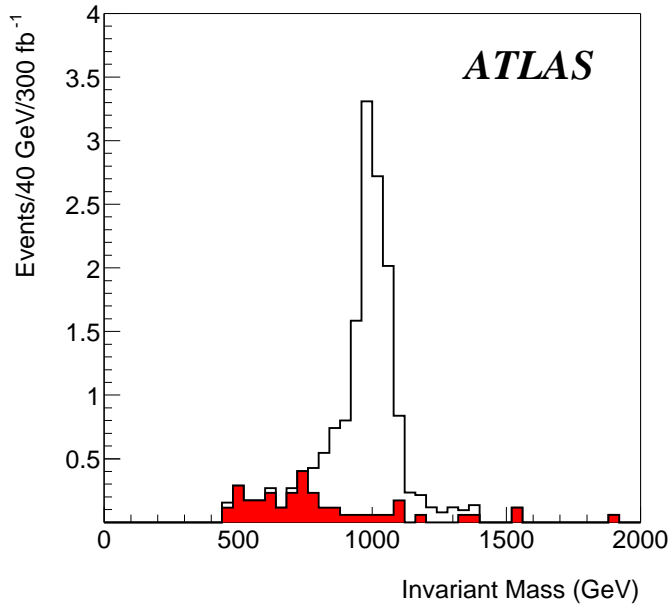}
\vskip2mm
\caption{Invariant mass of the $Zt$ pair, reconstructed from the $\ell^+\ell^-\ell^\pm \nu b$ final state. The signal (white) is $T\to Zt$, computed for $M_T=1$ TeV, $\lambda_1/\lambda_2=1$, and Br$(T\to Zt)=25$\%. The background (red) is dominated by $tbZ$. From Ref.~\cite{ATLAS}.}
\label{fig:ATLAS_top}
\end{center}
\end{figure}

All three SM decay modes in Eq.~\leqn{Tdecays} provide characteristic signatures for the discovery of the $T$ at the LHC. A detailed study of the LHC discovery potential in each decay mode has been preformed by the ATLAS collaboration~\cite{ATLAS}. The $Wb$ signal, reconstructed via the $\ell\nu b$ final state, was found to be the most promising, with the $5\sigma$ discovery reach of 2000 (2500) GeV for $\lambda_1/\lambda_2=1$ (2) and 300 fb$^{-1}$ integrated luminosity. The $Zt$ channel, reconstructed using leptonic $Z$ decays and $t\to Wb \to \ell\nu b$, provides a clean signature with small backgrounds, as shown in Fig.~\ref{fig:ATLAS_top}. However, the discovery reach is somewhat below that for the $Wb$ mode due to lower statistics: 
1050 (1400) GeV with $\lambda_1/\lambda_2=1$ (2) and 300 fb$^{-1}$. The $ht$ mode is more challenging, but if the $T$ quark is observed in other channels and its mass is known, the $ht$ signal can be separated from background and used to check the decay pattern in Eq.~\leqn{Tdecays}.

A discovery of the $T$ quark would not by itself provide a smoking gun signal for the LH model, since many competing theories predict particles with similar properties at the TeV scale. To prove that the LH interpretation of this particle is correct, it would be desirable to confirm its role in canceling the one-loop quadratic divergence by testing the relation between the $T$ mass and its couplings predicted by the LH models. Examples of such relations are provided by Eq.~\leqn{yuk_rel}, valid in the L$^2$H model, or Eq.~\leqn{simple_sumrule}, valid in the $SU(3)$ simple group model with $f_1=f_2$. In both examples, all four quantities\footnote{In other models, the corresponding relation may involve more 
parameters and testing it could require additional independent measurements. 
For example, an independent measurement of the ratio $f_1/f_2$ would be 
required to confirm the LH cancellation in a simple group model with $f_1\not=f_2$.} which enter these relations are in principle observable~\cite{PPP}: the top Yukawa $\lambda_t$ is already known, the $T$ mass will be directly measured at the LHC (see Fig.~\ref{fig:ATLAS_top}), and if the $W_H$ gauge bosons are discovered, a measurement of their mass and production cross section will determine the scale $f$. Measuring $\lambda_T$ will be more difficult: while in principle this quantity can be extracted from the single $T$ production cross section, the large uncertainty in the $b$ quark pdf would severely limit the precision of this measurement. Moreover, the total production cross section measurement requires that all $T$ decay channels be reconstructed, which may be challenging in practice. 

The LHC phenomenology of the $T$ quark in the $SU(3)$ simple group LH model has been studied in Ref.~\cite{Smoke}. The $T$ production cross sections are generically similar in size to those in the L$^2$H model, and its SM decays still have the widths given by Eq.~\leqn{Tdecays}. New $T$ decay channels, such as $t\eta$, $tY^0$ and $bX^+$, may be kinematically allowed, depending on the details of the spectrum. However, as long as the branching ratios into the SM modes are sizable, the same search strategies as in the Littlest Higgs model would apply. In addition to the heavy top, the simple group model contains three generations of vector-like TeV-scale quarks $U_H$ and leptons $N_H$. The LHC phenomenology of these particles has also been discussed in Ref.~\cite{Smoke}.

The top sector of the Littlest Higgs model with T parity contains two new vector-like quarks\footnote{In the modified T-parity symmetric embedding of the top sector suggested in Ref.~\cite{Tnew}, there is no T-even heavy top; the phenomenology of the T-odd $T$ state is very similar to the original model discussed here.}: the T-even state $T_+$ and the T-odd state $T_-$. Comparing Eqs.~\leqn{heavytopmass} and~\leqn{Tminusmass} shows that the T-odd state is always the lightest of the two, $M_{T_-} < M_{T_+}$. The collider phenomenology of the T-even state~\cite{JP} is very similar to that in a model with no T parity: single $T_+$ production is allowed and is the dominant channel for sufficiently high masses, and the decays into $Wb$, $tZ$ and $th$ final states occur with partial widths given in Eq.~\leqn{Tdecays}. The only new feature is that in certain regions of parameter space a new decay channel $T_+\to T_- B_H$ is allowed; followed by the decay $T_-\to tB_H$, this mode leads to a top+missing $E_T$ signature. The T-odd quarks need to be pair-produced; at the LHC, $q\bar{q}\to T_-\bar{T}_-$ and $gg\to T_-\bar{T}_-$ are the dominant production channels. The produced $T_-$ quarks decay into $tB_H$; if the $T_-$ is sufficiently heavy, other decay channels, such as $bW_H^+$, may be allowed. All events with $T_-$ production are expected to have large amount of missing transverse momentum due to the presence of two weakly interacting, stable $B_H$ bosons in each event~\cite{JP}. The LHT model also contains three generations of vector-like TeV-scale T-odd fermions, the partners of light quark and lepton doublets. The collider phenomenology of these states has not yet been studied in detail. It is clear, however, that pair-production of these states, followed by their cascade decays down to the LTP $B_H$, will result in events with high-$p_T$ jets and/or leptons and large missing transverse momentum. Such signatures are also expected in the R-parity conserving SUSY theories such as the MSSM, and in models with universal extra dimensions (UED) with conserved KK parity~\cite{ued,fooled}. This raises an interesting question of whether these models can be experimentally distinguished at the LHC and the ILC.

While discovering and studying the $T$ quark provides the most direct 
experimental window into the top sector of Little Higgs models, the models 
also predict potentially observable deviations in the gauge and Yukawa 
couplings of the familiar top quark $t$ from their SM values, due to the
mixing of $t$ and $T$. In particular,
Ref.~\cite{topSnow05} found that the expected precision of the 
$e^+e^-\to t\bar{t}$ cross section and the top width measurements at the ILC 
will be sufficient to observe the shifts in the $ttZ$ and $tbW$ 
couplings predicted by the Littlest Higgs model, throughout the model
parameter range preferred by naturalness.

\subsection{Scalar Sector}

The presence and the salient features of the TeV-scale gauge bosons and $T$ quarks are completely generic in the LH models, since these particles are required for the collective symmetry breaking mechanism to work. The scalar sector of the LH models, on the other hand, is more model-dependent: explicit models typically contain additional scalar states, but their spectrum and properties vary. For example, the Littlest Higgs model contains a weak triplet scalar field $\phi$. One of the components of the triplet, the doubly-charged scalar $\phi^{++}$, could provide interesting collider signatures, such as a resonant contribution to the $W^+W^+$ elastic scattering process~\cite{Han}.   
(In the Littlest Higgs model with T parity, the triplet is T-odd and $\phi^{++}$, along with other components, need to be pair-produced; the production cross section, decay channels and signatures have been studied in Ref.~\cite{JP}.) Many LH models contain electroweak-singlet pseudoscalar particles: examples include the Littlest Higgs model with an $SU(2)\times SU(2)\times U(1)_Y$ gauge group, where the $\eta$ state (see Eq.~\leqn{pions}) is not absorbed and remains physical, as well as the $SU(3)$ simple group model. The phenomenology of these states has been considered in Ref.~\cite{axion}; in particular, it was found that their inclusive production at the LHC may yield impressive diphoton resonances.

The L$^2$H model, along with many other Little Higgs models, contains a Higgs boson whose properties are identical to the SM Higgs in the limit $f\to\infty$. For finite $f$, however, the LH models predict deviations from the SM, at the level of $v^2/f^2$, in observables such as Higgs production and decay rates. Corrections at a few per cent level can be induced. For example, the Higgs decay rates in the Littlest Higgs model were analyzed in Ref.~\cite{Hdecays}, where it was found that, for $f=1$ TeV, $\Gamma(H \to gg)$ is reduced by 6--10\% and $\Gamma(H \to \gamma \gamma)$ is reduced by 5--7\% compared to their SM values. (The LH effects in the decay $H\to \gamma Z$ were considered in Ref.~\cite{gammaZ}.) A sensitive search for the predicted deviations of the Higgs coupling to photon pairs could be performed by the ILC running in the $\gamma\gamma$ mode, using the process $\gamma\gamma\to H\to b\bar{b}$~\cite{gammas}. 

\section{Other Topics}
\label{misc}

Despite their relative novelty, the literature on the Little Higgs models is quite extensive, and many aspects of these theories have been investigated. In this section, we will briefly review the results on several topics not covered above: detailed analyses of fine-tuning in the EWSB sector; ultraviolet completions of the LH models; flavor aspects; and predictions for cosmology.

\subsection{Fine-Tuning and EWSB}

\begin{figure}[t]
\begin{center}
\epsfig{file=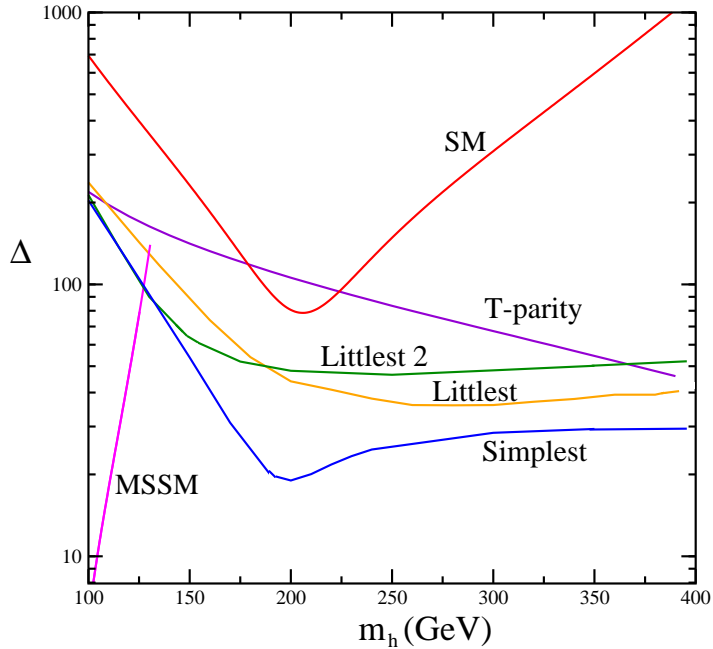,width=9.5cm,angle=-90}
\vskip2mm
\caption{The degree of fine-tuning $\Delta$ in the SM with a cutoff at 10 TeV; the MSSM; and four LH models. From Ref.~\cite{FT1}.}
\label{fig:FT}
\end{center}
\end{figure}

Simple estimates of the sizes of various terms in the CW potential of the LH models suggest that the Higgs vev of 246 GeV can be obtained with little or no fine-tuning, as long as the scale $f$ is around 1 TeV. A more detailed analysis of the required degree of fine-tuning was presented in Ref.~\cite{FT1}. While there is no unique way to quantify fine-tuning, some sensible measures can be defined, and at least the relative degree of tuning in various models can be compared if the same measure is applied to them. The authors of~\cite{FT1} use the measure pioneered by Barbieri and Giudice~\cite{BG} in the context of supersymmetric models: writing the Higgs vev as $v^2=v^2(p_1, p_2, \ldots)$, where $p_i$ are the input (Lagrangian) parameters of the model under study, the amount of fine tuning is given by
\beq
\Delta = \left[ \sum_i \Delta_i^2\right]^{1/2},
\eeq{ft1}
where $\Delta_i$ are defined via
\beq
\frac{\delta v^2}{v^2} = \Delta_i\,\frac{\delta p_i}{p_i}.
\eeq{ft2}
Four models, the original Littlest Higgs, the Littlest Higgs with a single gauged $U(1)$ (``Littlest 2''), the Littlest Higgs with T parity, and the $SU(3)$ simple group model, have been analyzed. The results are shown in Fig.~\ref{fig:FT}. In all cases, the authors conclude that the degree of fine-tuning is higher than the value suggested by naive estimates, as well as the values obtained in the
MSSM if the Higgs mass is very close to the current experimental lower bound, $m_h\gapproxeq 114$ GeV. It is interesting to note that in the L$^2$H model with T parity, fine-tuning decreases with the increase in $m_h$; values of $m_h$ as high as 800 GeV can be consistent with precision electroweak fits in this model (see Ref.~\cite{HMNP} and section~\ref{pewT}). Similarly,
the analysis of the exact one-loop CW potential in Ref.~\cite{FT2} has found that the natural expectation in the L$^2$H model, in the absence of fine-tuning, is that the Higgs is ``cruiserweight'', in the 800 GeV range for $f\sim 2$ TeV. 

\subsection{Ultraviolet Completions}

All LH models are effective field theories, with a cutoff scale around 10 TeV. Beyond that scale, a more fundamental theory, the ultraviolet (UV) completion of the LH model, is required. While the relevant energy scales will not be probed directly in collider experiments in the near future, it is important to construct and study explicit UV completions of LH models. Indeed, the constraints from low-energy experiments require that the four-fermion operators that generate flavor-changing neutral currents (FCNCs) not be generated at scales below $\sim 1000$ TeV, well above the regime of validity of the LH description. A successful LH model thus needs to be supplemented by a UV completion with a non-generic flavor structure 
which would satisfy this constraint. Also, the CW potential of the LH models receives contributions sensitive to the cutoff physics, see section~\ref{ewsb}. If a UV completion is specified, these contributions could be computed, resulting in a completely predictive model of EWSB\footnote{With an additional assumption 
of the ``vector limit''~\cite{vl}, the Higgs potential can be computed within 
the effective theory~\cite{technirho}.}.

Several examples of UV completions of popular LH models have been proposed. For instance, Ref.~\cite{UV1} embedded a slightly modified version of the Littlest Higgs model into a supersymmetric theory, which is renormalizable and valid up to scales of order $M_{\rm Pl}$. The theory contains a new ``ultracolor'' gauge interaction (the model of Ref.~\cite{UV1} takes the ultracolor gauge group to be $SO(7)$), and a set of ``ultrafermions'', which transform in a real representation of $SO(7)$. SUSY is broken softly at the 10 TeV scale, driving the theory into a confining phase; the $SU(5)\to SO(5)$ symmetry breaking pattern arises from condensation of ultrafermions. 
As an alternative, the $SU(5)/SO(5)$ Littlest Higgs model can be UV-completed by implementing it in a slice of five-dimensional Anti-de Sitter space~\cite{UVAdS}. In this setup, the $SU(5)$ global symmetry of the 4D L$^2$H model appears as a {\it gauged} $SU(5)$ symmetry in the 5D bulk, broken down to $SO(5)$ on the IR brane. Collective breaking appears in the choice of brane boundary conditions, rather than the gauge couplings.   

Refs.~\cite{turtle,LHtower} investigated the interesting possibility that the UV completion of an LH theory is provided by another LH theory, extending the perturbative description by a factor of 10, to $\sim 100$ TeV.  Explicit toy models implementing this idea were presented: for example, in Ref.~\cite{turtle} the $SU(3)$ simple group LH model was UV completed by two copies of the Littlest Higgs at higher energies. In principle, this procedure could be iterated to obtain a theory with light scalars and an exponentially large cutoff in the absence of supersymmetry or strong dynamics.  

\subsection{Flavor Physics}

Apart from the UV operator contributions mentioned above, flavor-violating low-energy observables are also generically expected to receive contributions from the weakly coupled TeV-scale new states predicted by the LH models. Several studies have evaluated these effects in the context of the L$^2$H model. For example, box diagrams involving $W_H^\pm$ gauge bosons and/or the heavy top quark $T$ contribute to the $K_0-\bar{K}_0$ and $B_{d,s}^0-\bar{B}_{d,s}^0$ mixing mass differences, as well as the CP-violating parameter $\varepsilon_K$~\cite{Buras1,Buras2} (see also~\cite{bantib}). The $K^0\to \pi^0\nu\bar{\nu}$~\cite{Kpinunubar} and $B\to s\gamma$~\cite{bsgamma} decay rates also receive contributions from the diagrams with these states running in the loop. Because of the presence of the heavy top $T$, flavor-changing neutral currents are also induced at tree level in the up sector; the induced rates of rare $D$ meson decays, $t\to cZ$, and the $D_0-\bar{D}_0$ mass difference have been considered in Ref.~\cite{Lee}. Unfortunately, in all cases, the experimental sensitivity and the theoretical uncertainty inherent in hadronic observables imply that the bounds on the model parameters from the non-observation of the predicted flavor-changing effects are significantly weaker than the precision electroweak constraints.

Neutrino masses and lepton-flavor violation in the L$^2$H model have been considered in Ref.~\cite{numass}. It was argued that the most satisfactory way of incorporating neutrino masses is to include a lepton-number violating interaction between the scalar triplet and lepton doublets. In the absence of T parity, the triplet acquires a vev, inducing a Majorana mass for $\nu_L$; however, there is no see-saw mechanism, and either the vev or the coupling need to be fine-tuned to obtain the masses at the experimentally observed scale. Potential collider signatures of this scenario, arising from the lepton-number violating decays of triplet scalars, were also considered. Ref.~\cite{numassnat} suggested a Little Higgs model, based on a modification of the simple group $SU(3)$ theory, in which the observed neutrino masses and mixings can be accommodated naturally, without fine-tuning. This model contains three additional Dirac neutrinos at the TeV scale. (Neutrino masses in the simple group LH model have also been considered in Ref.~\cite{Lee2}, while lepton-flavor violating effects from a possible UV completion of the L$^2$H model were studied in Ref.~\cite{Lee3}.)

\subsection{Little Higgs Cosmology}

\begin{figure}[t]
\begin{center}
\includegraphics[width=9cm]{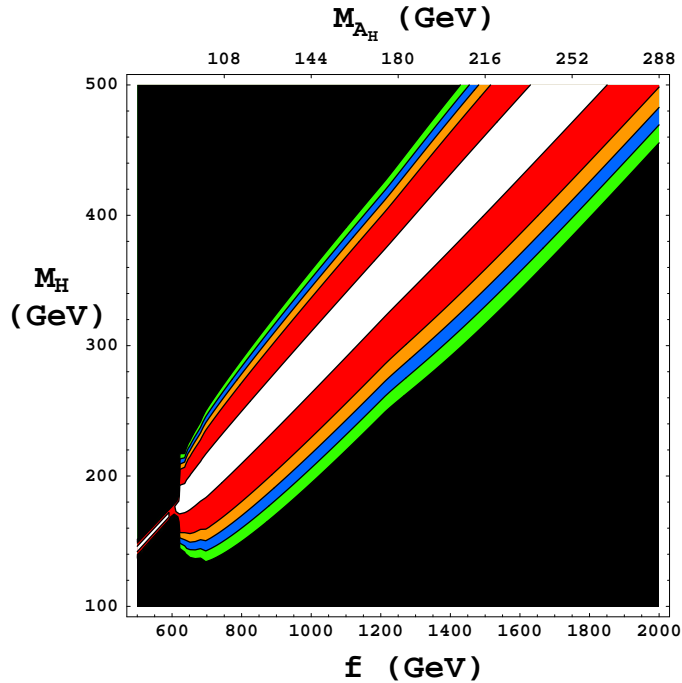}
\vskip2mm
\caption{Contours of constant relic density of the ``heavy photon'' in the L$^2$HT model. In order from lightest to darkest regions, the heavy photon makes up 0-10\%, 10-50\%, 50-70\%, 70-100\%, 100\%, and $>100$\% of the dark matter density measured by WMAP~\cite{WMAP}. From Ref.~\cite{JP}.}
\label{fig:LHDM}
\end{center}
\end{figure}

As with any model of EWSB, LH models have interesting consequences for early universe cosmology, especially the electroweak phase transition era. The dynamics of the transition were analyzed in Ref.~\cite{ELR}. Explicit calculations were performed in the context of the L$^2$H model with a single gauged $U(1)$, but the main features uncovered by the analysis are expected to be generic. Surprisingly, it was found that the electroweak symmetry in this model is {\it not} restored at high temperatures. Rather, in the high-temperature limit the thermal CW potential develops a symmetry-breaking minimum with an order parameter of order $f$. Another surprising result, with possible relevance for cosmology, is the rich spectrum of allowed topological defects in the L$^2$H model~\cite{Trodden}.

Some LH models contain discrete symmetries that could render one of the new TeV-scale particles stable. If this particle is weakly interacting, it can provide an attractive dark matter candidate. Two examples of LH dark matter candidates have been studied in the literature: one arising from a theory space model~\cite{AnJay}, and another one from the L$^2$H model with T parity~\cite{JP}. For example, the relic abundance of the WIMP candidate of the L$^2$HT model, the ``heavy photon'' (or, more precisely, the heavy partner of the hypercharge gauge boson), is shown in Fig.~\ref{fig:LHDM}. The dominant WIMP annihilation channels in this model are $W^+W^-$ and $ZZ$, via an $s$-channel Higgs exchange. Note that the WMAP measurement of the dark matter abundance~\cite{WMAP} provides a strong constraint on the parameter space of the model. Direct and indirect signatures of the heavy photon dark matter are currently under investigation~\cite{BNPS}.

\section{Conclusions}
\label{conc}

In this article, we reviewed the Little Higgs models of electroweak symmetry 
breaking. In these models, the Higgs boson is a composite particle. While a
generic theory with a composite Higgs becomes strongly coupled around the 
TeV scale, leading to phenomenological difficulties, the LH models remain 
perturbative up to the 10 TeV scale. Two ingredients are required to 
stabilize the ``little hierarchy'' between the Higgs mass and the strong 
coupling scale. First, the Higgs is identified with a Nambu-Goldstone field
corresponding to a dynamically broken global symmetry; this guarantees that
its mass vanishes at tree level. Second, while the global symmetry is also 
broken explicitly in order to describe the gauge and Yukawa interactions of 
the Higgs, the explicit breaking is ``collective''. That is, none of the 
coupling constants in the LH Largangian by itself breaks all the global
symmetries required to keep the Higgs massless; it is only their collective
effect that results in the Higgs acquiring a mass and non-derivative 
interactions. This collective symmetry breaking mechanism guarantees the
absence of quadratically divergent one-loop contributions to the Higgs
mass. While a quadratic divergence reappears at the two loop order, and
logarithmically divergent contributions to the Higgs mass are induced at
one loop, these contributions are sufficiently small and no fine tuning is
required to stabilize the little hierarchy. At the same time, the 
logarithmically divergent one-loop contibution to the Higgs mass from top
loops, which typically dominates over other effects due to a large value of
the top Yukawa coupling in the SM, has the correct sign to trigger EWSB, and
an order-one Higgs quartic coupling is generated at one loop. Thus, the
LH models provide a phenomenologically consistent, simple and attractive 
picture of radiatively induced EWSB.
 
A large number of explicit, fully realistic models of EWSB implementing the 
collective symmetry breaking mechanism have been constructed; several of
them have been reviewed in this article. All models predict new vector-like
fermions and gauge bosons at the TeV scale; these are required to cancel the
one-loop quadratic divergences in the Higgs mass. Many specific models also
contain additional scalar states. The new physics at the TeV scale induces
corrections to weak scale observables, many of which have been measured 
with exquisite precision by LEP, SLD, and other experiments. The simplest LH 
models, such as the Littlest Higgs and the $SU(3)$ simple group model, turn 
out to be significantly constraied by precision electroweak measurements, 
although they may be consistent in some special regions of the parameter space.
An interesting solution to this difficulty is provided by the models with 
T parity, in which the weak-scale observables receive no tree-level 
corrections, and a consistent fit to data is possible for a wide range of 
model parameters.

The new particles predicted by the LH models should be within the discovery 
range of the LHC experiments. The suggested signatures, estimates of
experimental sensitivities, and possible experimental strategies for 
confirming the LH origin of the observed new particles, were reviewed in 
this article. Much work remains to be done, however, to maximize the
experiments' efficiency in searching for the LH signatures. This includes 
systematically incorporating the LH models in the existing Monte Carlo
generators, which is required for systematic comparison of the model
predictions with the LHC data. Also, more work needs to be done on the 
colider phenomenology of models with T parity, whose signatures can ``fake''
those expected in the simple SUSY theories such as the MSSM.

Several interesting model-building issues related to the LH models remain to 
be addressed. For example, detailed studies of the Higgs effective potential
in a few representative LH models indicate that the models are somewhat more 
fine-tuned than suggested by naive estimates, especially if the Higgs is 
light. It is not clear whether this is a generic phenomenon, and it would be 
interesting if explicit counterexamples could be constructed. Also, while
several UV completions of the LH models (describing the physics above the
10 TeV scale) have been built, the existing proposals are rather complicated.
It would be interesting to see if the LH theories could emerge from a simpler 
UV dynamics.

While we attempted to provide a comprehensive review of the existing LH models 
of EWSB, several closely related subjects remained beyond the scope of 
this article. These include the recently proposed alternative implementation
of the Higgs as a pseudo-NGB, stabilized by a ``mirror'' symmetry 
interchanging the SM particles with their mirror 
partners~\cite{twin,twinpheno}. The ``intermediate Higgs'' 
models~\cite{intermediate} implement the collective symmetry breaking 
mechanism in the top sector, but not in the gauge sector, so that the one-loop 
quadratic divergences from the bow tie gauge loops are uncanceled. Some 
amount of fine-tuning is then required to stabilize the little hierarchy or
to avoid the constraints from operators such as~\leqn{ops3}. However, the
models are very simple, and the perturbatively calculable corrections to 
precision electroweak observables are small (loop-level only). Another 
interesting recent development is a series of papers which
attempt to implement the collective symmetry breaking mechanism in models 
with low-energy supersymmetry, in order to reduce the amount of 
fine-tuning required for successful EWSB in simple SUSY models such as 
the MSSM~\cite{susy1,susy2,susy3,susy4,susy5}.  
The collective symmetry breaking mechanism has been applied in other contexts where scalar masses need to be stabilized against radiative corrections: examples include the light flavon fields of Refs.~\cite{flavon}, as well as radiatively stable quintessence~\cite{quint} and inflaton~\cite{inflate} fields. 
The interested reader is encouraged to consult the original papers for a 
detailed discussion of these ideas.

\vskip.5cm
\section*{Acknowledgements}

Over the years, I had numerous illuminating discussions with many colleagues
on the topics related to the subject of this review. I am especially grateful 
to my collaborators on papers exploring aspects of Little Higgs models and 
their phenomenology: Gustavo Burdman, Jay Hubisz, Patrick Meade, Andrew Noble,
Michael Peskin, and Aaron Pierce. I would also like to thank Nima 
Arkani-Hamed, Csaba Csaki, Hsin-Chia Cheng, Thomas Gregoire, Tao Han,
Heather Logan, Yasunori Nomura, Frank Petriello, Martin Schmaltz, Tim Tait, 
and Jay Wacker for useful discussions. I am indebted to Patrick 
Meade and Andrew Noble for reading the manuscript of this article before it 
was submitted, and making suggestions that led to significant improvements.
Needless to say, I alone am responsible for any mistakes or omissions that
may remain in this version of the article.

My research is supported by the National Science Foundation under grant 
PHY-0355005.


\if

In this subsection, some details from one of the studies of precision electroweak corrections in a Littlest Higgs model (Ref.~\cite{PPP}) will be presented, with the goal of providing an explicit example of the calculations involved.

 with $SU(2)\times SU(2) \times U(1)$ gauge

In terms of these eigenstates, Eq.~\leqn{topyuk} has the form
\beq
{\cal L}^T_{\rm m} =
\, \lambda_1 f \left[ \half(1+c_v) \bar{U}_{L+} + \frac{s_v}{\sqrt{2}} \bar{u}_{L+} \right] u_R
+ \lambda_2 f \left( \bar{U}_{L+} U_{R+} + \bar{U}_{L-} U_{R-} \right) +
{\rm h.c.}
\eeq{topmass}
where we have used Eq.~\leqn{sigmafull}. The T-odd
states $U_{L-}$ and $U_{R-}$ combine to form a Dirac fermion $T_-$,
with mass $m_{T_-}=\lambda_2 f$. The remaining T-odd states $q_-$ receive a Dirac mass from the interaction in Eq.~\leqn{heavyyuk}, and are assumed to be decoupled. The mass terms for the T-even states are diagonalized by defining
\beqa
t_L &=& \cos\beta \,u_{L+} - \sin\beta \,U_{L+},~~~~~~~
T_{L+} = \sin\beta \,u_{L+} +\cos\beta \,U_{L+},\CR
t_R &=& \cos\alpha \,u_R - \sin\alpha \,U_{R+},~~~~~~~
T_{R+} = \sin\alpha \,u_R + \cos\alpha \,U_{R+},
\eeqa{rot2}
where $t$ is identified with the SM top and $T_+$ is its T-even heavy partner. The mixing angles are given by
\beqa
\alpha &=& \half \tan^{-1} \frac{4\lambda_1\lambda_2 (1+c_v)}{4\lambda_2^2-\lambda_1^2(2s_v^2+(1+c_v)^2)}\, ,\CR
\beta &=& \half \tan^{-1} \frac{2\sqrt{2}\lambda_1^2 s_v (1+c_v)}{4\lambda_2^2+(1+c_v)^2\lambda_1^2-2\lambda_1^2 s_v}.
\eeqa{mixingsexact}
To leading order in the $v/f$ expansion,
\beq
\sin\alpha \,=\, \frac{\lambda_1}{\sqrt{\lambda_1^2+\lambda_2^2}},~~~
\sin\beta \,=\, \frac{\lambda_1^2}{\lambda_1^2+\lambda_2^2}\,\frac{v}{f}.
\eeq{mixings}
The masses of the two T-even Dirac fermions are given by
\beq
m^2_{t,T_+} \,=\,f^2\Delta\,\left(1\pm \sqrt{1-\frac{\lambda_1^2\lambda_2^2s_v^2}{2\Delta^2}}\right),
\eeq{massesexact}
where
\beq
\Delta \,=\, \half\,\left(\lambda_2^2+\frac{\lambda_1^2}{2}(s_v^2+\half(1+c_v)^2)
\right).
\eeq{Delta}
To leading order in $v/f$, 
\beq
m_t=\frac{ \lambda_1 \lambda_2
v}{\sqrt{\lambda_1^2+\lambda_2^2}},~~~~\mT = \sqrt{\lambda_1^2+\lambda_2^2}\, f.
\eeq{tmass}
 It is interesting to note that the T-odd states do not participate in the cancellation of quadratic divergences in the top sector: the cancellation only involves loops of $t$ and $T_+$, and the details are identical to the LH model without T parity~\cite{PPP}.

Using the above equations, it is straightforward to obtain the Feynman rules for the top sector of the LH model; we list the rules relevant for the calculations in this paper in Table~\ref{tab:Feynman}.
\fi

\if
The spectrum of the model can be straightforwardly obtained using the following formula for the vev of the $\Sigma$ field, including the EWSB effects:
\beq
\left< \Sigma \right> \,=\, \left(\begin{array}{ccccc}
0& 0& 0& 1& 0\\
0& -\half(1-c_v)& \frac{i}{\sqrt{2}}s_v& 0& \half(1+c_v)\\
0& \frac{i}{\sqrt{2}}s_v& c_v& 0& \frac{i}{\sqrt{2}}s_v\\
1& 0& 0& 0& 0\\
0& \half(1+c_v)& \frac{i}{\sqrt{2}}s_v& 0& -\half(1-c_v)\\
\end{array}\right),
\eeq{sigmafull}
where
\beq
s_v=\sin\frac{\sqrt{2}v}{f},~~~c_v=\cos\frac{\sqrt{2}v}{f}.
\eeq{sincos}
The charged gauge boson matrix is given by
\beq
    M_+^2 =  {f^2\over 4} \pmatrix{ g_1^2     &    -  \half g_1g_2 (1+c_v) \cr
                                          -\half g_1g_2 (1+c_v) &    g_2^2\cr}.
\eeq{firstmW}
Diagonalizing this matrix yields the mass eigenstates,
\beqa
W_L^\pm &=& \cos \theta_+ W_1^\pm + \sin \theta_+ W_2^\pm,\CR
W_H^\pm &=& -\sin \theta_+ W_1^\pm + \cos \theta_+ W_2^\pm,  
\eeqa{Wmass}
where the mixing angle is given by
\beq
\theta_+\,=\,\frac{1}{2}\tan^{-1} \frac{g_1g_2(1+c_v)}{g_2^2-g_1^2}.
\eeq{Wmixing}
To leading order in the $v/f$ expansion, $\theta_+=\psi$. The masses of the two eigenstates are given by
\beq
M^2_{L/H} \,=\,\frac{f^2}{8}\,\left(g_1^2+g_2^2\,\pm\,\frac{g_1^2-g_2^2}{\cos 2\theta_+}\right).
\eeq{Wmassexact}
To leading order in $v/f$, these formulas reduce to
\beq
M^2_L = \frac{g^2v^2}{4},~~~M^2_H = (g_1^2+g_2^2)\,\frac{f^2}{4}\,=\, \frac{g^2 f^2}{\sin^2 2\psi}.
\eeq{Wmassapprox}

\draftnote{[Formulas for the neutral gauge boson masses and mixings go in here -  leading order in $v/f$ expansion?]}
\fi


\begin{thebibliography}{99}

{\small

\bibitem{PDG}
 S.~Eidelman {\it et al.}  [Particle Data Group],
  Phys.\ Lett.\ B {\bf 592}, 1 (2004).


\bibitem{c1}
 D.~B.~Kaplan and H.~Georgi,
  Phys.\ Lett.\ B {\bf 136}, 183 (1984).

\bibitem{c2}
  D.~B.~Kaplan, H.~Georgi and S.~Dimopoulos,
  Phys.\ Lett.\ B {\bf 136}, 187 (1984).

\bibitem{c3}
  H.~Georgi, D.~B.~Kaplan and P.~Galison,
  Phys.\ Lett.\ B {\bf 143}, 152 (1984).

\bibitem{c4}
  H.~Georgi and D.~B.~Kaplan,
  Phys.\ Lett.\ B {\bf 145}, 216 (1984).

\bibitem{c5}
  M.~J.~Dugan, H.~Georgi and D.~B.~Kaplan,
  Nucl.\ Phys.\ B {\bf 254}, 299 (1985).

\bibitem{decon}
  N.~Arkani-Hamed, A.~G.~Cohen and H.~Georgi,
  Phys.\ Rev.\ Lett.\  {\bf 86}, 4757 (2001)
  [arXiv:hep-th/0104005].

\bibitem{decon1}
  C.~T.~Hill, S.~Pokorski and J.~Wang,
  Phys.\ Rev.\ D {\bf 64}, 105005 (2001)
  [arXiv:hep-th/0104035].

\bibitem{A5}
  N.~S.~Manton,
  Nucl.\ Phys.\ B {\bf 158}, 141 (1979);
  Y.~Hosotani,
  Phys.\ Lett.\ B {\bf 126}, 309 (1983).

  
\bibitem{bigmoose}
  N.~Arkani-Hamed, A.~G.~Cohen and H.~Georgi,
  Phys.\ Lett.\ B {\bf 513}, 232 (2001)
  [arXiv:hep-ph/0105239].


\bibitem{review}
 M.~Schmaltz and D.~Tucker-Smith,
  arXiv:hep-ph/0502182.


\bibitem{LH}
  N.~Arkani-Hamed, A.~G.~Cohen, E.~Katz and A.~E.~Nelson,
  JHEP {\bf 0207}, 034 (2002)
  [arXiv:hep-ph/0206021].

\bibitem{Georgibook}
  H.~Georgi,
{\it Weak Interactions And Modern Particle Theory}, 
Menlo Park: Benjamin/Cummings, 1984. \\
See also {\tt http://schwinger.harvard.edu/\~{ }georgi/283.html}

\bibitem{BPRS}
  R.~Barbieri, A.~Pomarol, R.~Rattazzi and A.~Strumia,
  Nucl.\ Phys.\ B {\bf 703}, 127 (2004)
  [arXiv:hep-ph/0405040].

\bibitem{HS}
  Z.~Han and W.~Skiba,
  Phys.\ Rev.\ D {\bf 71}, 075009 (2005)
  [arXiv:hep-ph/0412166].

\bibitem{CW}
 S.~R.~Coleman and E.~Weinberg,
  Phys.\ Rev.\ D {\bf 7}, 1888 (1973).

\bibitem{BPP}
  G.~Burdman, M.~Perelstein and A.~Pierce,
  Phys.\ Rev.\ Lett.\  {\bf 90}, 241802 (2003)
  [Erratum-ibid.\  {\bf 92}, 049903 (2004)]
  [arXiv:hep-ph/0212228].
 
 \bibitem{Jarry}
A. Jarry, {\it Ubu Roi}, Paris: Mercure de France, 1896.
 
 \bibitem{PPP}
  M.~Perelstein, M.~E.~Peskin and A.~Pierce,
  Phys.\ Rev.\ D {\bf 69}, 075002 (2004)
  [arXiv:hep-ph/0310039].
 
\bibitem{CC2} 
 C.~Csaki, J.~Hubisz, G.~D.~Kribs, P.~Meade and J.~Terning,
  Phys.\ Rev.\ D {\bf 68}, 035009 (2003)
  [arXiv:hep-ph/0303236].
 
\bibitem{HMNP}
  J.~Hubisz, P.~Meade, A.~Noble and M.~Perelstein,
  arXiv:hep-ph/0506042.

\bibitem{cust}
P.~Sikivie, L.~Susskind, M.~B.~Voloshin and V.~I.~Zakharov,
Nucl.\ Phys.\ B {\bf 173}, 189 (1980).


\bibitem{Simple0}
   D.~E.~Kaplan and M.~Schmaltz,
  JHEP {\bf 0310}, 039 (2003)
  [arXiv:hep-ph/0302049].

 \bibitem{Smoke}
  T.~Han, H.~E.~Logan and L.~T.~Wang,
  arXiv:hep-ph/0506313.


\bibitem{LHT0}
  H.~C.~Cheng and I.~Low,
  JHEP {\bf 0309}, 051 (2003)
  [arXiv:hep-ph/0308199].

\bibitem{minmoose}
 N.~Arkani-Hamed, A.~G.~Cohen, E.~Katz, A.~E.~Nelson, T.~Gregoire and J.~G.~Wacker,
  JHEP {\bf 0208}, 021 (2002)
  [arXiv:hep-ph/0206020].

\bibitem{moose}
  H.~Georgi,
  Nucl.\ Phys.\ B {\bf 266}, 274 (1986).

\bibitem{genmoose}
  T.~Gregoire and J.~G.~Wacker,
  JHEP {\bf 0208}, 019 (2002)
  [arXiv:hep-ph/0206023].

\bibitem{moosepheno}
  N.~Arkani-Hamed, A.~G.~Cohen, T.~Gregoire and J.~G.~Wacker,
  JHEP {\bf 0208}, 020 (2002)
  [arXiv:hep-ph/0202089].

\bibitem{antisym}
  I.~Low, W.~Skiba and D.~Smith,
  Phys.\ Rev.\ D {\bf 66}, 072001 (2002)
  [arXiv:hep-ph/0207243].

\bibitem{custodial}
S.~Chang and J.~G.~Wacker,
  Phys.\ Rev.\ D {\bf 69}, 035002 (2004)
  [arXiv:hep-ph/0303001].

\bibitem{custodial_lh}
 S.~Chang,
  JHEP {\bf 0312}, 057 (2003)
  [arXiv:hep-ph/0306034].


\bibitem{LHT}
H.~C.~Cheng and I.~Low,
  JHEP {\bf 0408}, 061 (2004)
  [arXiv:hep-ph/0405243].
 
 \bibitem{LHT1}
 I.~Low,
  JHEP {\bf 0410}, 067 (2004)
  [arXiv:hep-ph/0409025].


\bibitem{JP}
  J.~Hubisz and P.~Meade,
  Phys.\ Rev.\ D {\bf 71}, 035016 (2005)
  [arXiv:hep-ph/0411264].

\bibitem{Tnew}
  H.~C.~Cheng, I.~Low and L.~T.~Wang,
  arXiv:hep-ph/0510225.


\bibitem{Simple1}
  M.~Schmaltz,
  JHEP {\bf 0408}, 056 (2004)
  [arXiv:hep-ph/0407143].

\bibitem{Kong}
 O.~C.~W.~Kong,
  J.\ Korean Phys.\ Soc.\  {\bf 45}, S404 (2004)
  [arXiv:hep-ph/0312060].

\bibitem{Simple2}
  W.~Skiba and J.~Terning,
  Phys.\ Rev.\ D {\bf 68}, 075001 (2003)
  [arXiv:hep-ph/0305302].


\bibitem{CC1}
 C.~Csaki, J.~Hubisz, G.~D.~Kribs, P.~Meade and J.~Terning,
  Phys.\ Rev.\ D {\bf 67}, 115002 (2003)
  [arXiv:hep-ph/0211124].
  
\bibitem{HPR}
  J.~L.~Hewett, F.~J.~Petriello and T.~G.~Rizzo,
  JHEP {\bf 0310}, 062 (2003)
  [arXiv:hep-ph/0211218].
  
\bibitem{Han}
  T.~Han, H.~E.~Logan, B.~McElrath and L.~T.~Wang,
  Phys.\ Rev.\ D {\bf 67}, 095004 (2003)
  [arXiv:hep-ph/0301040].

\bibitem{CD}
 M.~C.~Chen and S.~Dawson,
  Phys.\ Rev.\ D {\bf 70}, 015003 (2004)
  [arXiv:hep-ph/0311032].
  
\bibitem{Kilian}
  W.~Kilian and J.~Reuter,
  Phys.\ Rev.\ D {\bf 70}, 015004 (2004)
  [arXiv:hep-ph/0311095].


\bibitem{Marandella}
  G.~Marandella, C.~Schappacher and A.~Strumia,
  Phys.\ Rev.\ D {\bf 72}, 035014 (2005)
  [arXiv:hep-ph/0502096].
  
\bibitem{Skiba}
  Z.~Han and W.~Skiba,
  Phys.\ Rev.\ D {\bf 72}, 035005 (2005)
  [arXiv:hep-ph/0506206].
 
 
\bibitem{zbbar}
  C.~x.~Yue and W.~Wang,
  Nucl.\ Phys.\ B {\bf 683}, 48 (2004)
  [arXiv:hep-ph/0401214].
 
\bibitem{g-2}
  S.~C.~Park and J.~h.~Song,
  arXiv:hep-ph/0306112.
 
\bibitem{private}
N.~Arkani-Hamed and J.~Wacker, private communication.

\bibitem{He}
  S.~Chang and H.~J.~He,
  Phys.\ Lett.\ B {\bf 586}, 95 (2004)
  [arXiv:hep-ph/0311177].

\bibitem{PEW66}
  T.~Gregoire, D.~R.~Smith and J.~G.~Wacker,
  Phys.\ Rev.\ D {\bf 69}, 115008 (2004)
  [arXiv:hep-ph/0305275].
  
 \bibitem{PEWmin}
  C.~Kilic and R.~Mahbubani,
  JHEP {\bf 0407}, 013 (2004)
  [arXiv:hep-ph/0312053].

  \bibitem{custPEW}
  R.~Casalbuoni, A.~Deandrea and M.~Oertel,
  JHEP {\bf 0402}, 032 (2004)
  [arXiv:hep-ph/0311038].
 

\bibitem{PT}
  M.~E.~Peskin and T.~Takeuchi,
  Phys.\ Rev.\ D {\bf 46}, 381 (1992).

 


\bibitem{atlas}
``ATLAS detector and physics performance. Technical design report.  Vol. 2,''
CERN-LHCC-99-15.

\bibitem{ATLAS}
  G.~Azuelos {\it et al.},
  Eur.\ Phys.\ J.\ C {\bf 39S2}, 13 (2005)
  [arXiv:hep-ph/0402037].

\bibitem{ILC_BHdirect}
  G.~C.~Cho and A.~Omote,
  Phys.\ Rev.\ D {\bf 70}, 057701 (2004)
  [arXiv:hep-ph/0408099].
  
  \bibitem{egamma}
  C.~x.~W.~Yue and W.~Wang,
  Phys.\ Rev.\ D {\bf 71}, 015002 (2005)
  [arXiv:hep-ph/0411266].
  
\bibitem{ILC1}
  J.~A.~Conley, J.~Hewett and M.~P.~Le,
  arXiv:hep-ph/0507198.

\bibitem{Park}
  S.~C.~Park and J.~Song,
  Phys.\ Rev.\ D {\bf 69}, 115010 (2004).

\bibitem{ILCttbar}
  C.~X.~Yue, L.~Wang and J.~X.~Chen,
  Eur.\ Phys.\ J.\ C {\bf 40}, 251 (2005)
  [arXiv:hep-ph/0501186].
  
\bibitem{ILC_ZH}
  C.~x.~Yue, S.~z.~Wang and D.~q.~Yu,
  Phys.\ Rev.\ D {\bf 68}, 115004 (2003)
  [arXiv:hep-ph/0309113].

\bibitem{ued}
  T.~Appelquist, H.~C.~Cheng and B.~A.~Dobrescu,
  Phys.\ Rev.\ D {\bf 64}, 035002 (2001)
  [arXiv:hep-ph/0012100].

\bibitem{fooled}
  H.~C.~Cheng, K.~T.~Matchev and M.~Schmaltz,
  Phys.\ Rev.\ D {\bf 66}, 056006 (2002)
  [arXiv:hep-ph/0205314].

\bibitem{topSnow05}
  C.~F.~Berger, M.~Perelstein and F.~Petriello,
  arXiv:hep-ph/0512053.

\bibitem{axion}
  W.~Kilian, D.~Rainwater and J.~Reuter,
  Phys.\ Rev.\ D {\bf 71}, 015008 (2005)
  [arXiv:hep-ph/0411213].
     
\bibitem{Hdecays}
  T.~Han, H.~E.~Logan, B.~McElrath and L.~T.~Wang,
  Phys.\ Lett.\ B {\bf 563}, 191 (2003)
  [Erratum-ibid.\ B {\bf 603}, 257 (2004)]
  [arXiv:hep-ph/0302188].
  
\bibitem{gammaZ}
  G.~A.~Gonzalez-Sprinberg, R.~Martinez and J.~A.~Rodriguez,
  Phys.\ Rev.\ D {\bf 71}, 035003 (2005)
  [arXiv:hep-ph/0406178].
  
\bibitem{gammas}
  H.~E.~Logan,
  Phys.\ Rev.\ D {\bf 70}, 115003 (2004)
  [arXiv:hep-ph/0405072].
  
  
  
\bibitem{FT1}
  J.~A.~Casas, J.~R.~Espinosa and I.~Hidalgo,
  JHEP {\bf 0503}, 038 (2005)
  [arXiv:hep-ph/0502066].

\bibitem{BG}
  R.~Barbieri and G.~F.~Giudice,
  Nucl.\ Phys.\ B {\bf 306}, 63 (1988).

\bibitem{FT2}
  F.~Bazzocchi, M.~Fabbrichesi and M.~Piai,
  arXiv:hep-ph/0506175.
 
\bibitem{vl}
  H.~Georgi,
  Nucl.\ Phys.\ B {\bf 331}, 311 (1990).

\bibitem{technirho}
  M.~Piai, A.~Pierce and J.~Wacker,
  arXiv:hep-ph/0405242.

\bibitem{UV1}
  E.~Katz, J.~y.~Lee, A.~E.~Nelson and D.~G.~E.~Walker,
  arXiv:hep-ph/0312287.
 
\bibitem{UVAdS}
  J.~Thaler and I.~Yavin,
  JHEP {\bf 0508}, 022 (2005)
  [arXiv:hep-ph/0501036].
  
\bibitem{turtle}
  D.~E.~Kaplan, M.~Schmaltz and W.~Skiba,
  Phys.\ Rev.\ D {\bf 70}, 075009 (2004)
  [arXiv:hep-ph/0405257].
  
\bibitem{LHtower}
  P.~Batra and D.~E.~Kaplan,
  JHEP {\bf 0503}, 028 (2005)
  [arXiv:hep-ph/0412267].
  
    
 \bibitem{Buras1}
  A.~J.~Buras, A.~Poschenrieder and S.~Uhlig,
  Nucl.\ Phys.\ B {\bf 716}, 173 (2005)
  [arXiv:hep-ph/0410309].

\bibitem{Buras2}
  A.~J.~Buras, A.~Poschenrieder and S.~Uhlig,
  arXiv:hep-ph/0501230.

\bibitem{bantib}
  S.~R.~Choudhury, N.~Gaur, A.~Goyal and N.~Mahajan,
  Phys.\ Lett.\ B {\bf 601}, 164 (2004)
  [arXiv:hep-ph/0407050].

\bibitem{Kpinunubar}
  S.~R.~Choudhury, N.~Gaur, G.~C.~Joshi and B.~H.~J.~McKellar,
  arXiv:hep-ph/0408125.
  
\bibitem{bsgamma}
  W.~j.~Huo and S.~h.~Zhu,
  Phys.\ Rev.\ D {\bf 68}, 097301 (2003)
  [arXiv:hep-ph/0306029].
    
\bibitem{Lee}
  J.~Y.~Lee,
  JHEP {\bf 0412}, 065 (2004)
  [arXiv:hep-ph/0408362].
    

\bibitem{numass}
  T.~Han, H.~E.~Logan, B.~Mukhopadhyaya and R.~Srikanth,
  arXiv:hep-ph/0505260.
  
\bibitem{numassnat}
  F.~del Aguila, M.~Masip and J.~L.~Padilla,
  arXiv:hep-ph/0506063.
  
\bibitem{Lee2}
  J.~Y.~Lee,
  arXiv:hep-ph/0504136.
  
\bibitem{Lee3}
 J.~Y.~Lee,
  JHEP {\bf 0506}, 060 (2005)
  [arXiv:hep-ph/0501118].
    
    
  
   
\bibitem{ELR}
  J.~R.~Espinosa, M.~Losada and A.~Riotto,
  Phys.\ Rev.\ D {\bf 72}, 043520 (2005)
  [arXiv:hep-ph/0409070].
   
\bibitem{Trodden}
  M.~Trodden and T.~Vachaspati,
  Phys.\ Rev.\ D {\bf 70}, 065008 (2004)
  [arXiv:hep-ph/0404105].
  
\bibitem{AnJay}
  A.~Birkedal-Hansen and J.~G.~Wacker,
  Phys.\ Rev.\ D {\bf 69}, 065022 (2004)
  [arXiv:hep-ph/0306161].
  
\bibitem{WMAP} 
D.~N.~Spergel {\it et al.},
Astrophys.\ J.\ Suppl.\  {\bf 148}, 175 (2003)
[astro-ph/0302209].
  
\bibitem{BNPS}
A.~Birkedal, A.~Noble, M.~Perelstein and A.~Spray, in preparation.  

\bibitem{twin}
  Z.~Chacko, H.~S.~Goh and R.~Harnik,
  arXiv:hep-ph/0506256.

\bibitem{twinpheno}
  R.~Barbieri, T.~Gregoire and L.~J.~Hall,
  arXiv:hep-ph/0509242.

\bibitem{intermediate}
  E.~Katz, A.~E.~Nelson and D.~G.~E.~Walker,
  JHEP {\bf 0508}, 074 (2005)
  [arXiv:hep-ph/0504252].


\bibitem{susy1}
  A.~Birkedal, Z.~Chacko and M.~K.~Gaillard,
  JHEP {\bf 0410}, 036 (2004)
  [arXiv:hep-ph/0404197].
  
\bibitem{susy2}
  P.~H.~Chankowski, A.~Falkowski, S.~Pokorski and J.~Wagner,
  Phys.\ Lett.\ B {\bf 598}, 252 (2004)
  [arXiv:hep-ph/0407242].
  
\bibitem{susy3}
  Z.~Berezhiani, P.~H.~Chankowski, A.~Falkowski and S.~Pokorski,
  arXiv:hep-ph/0509311.
  
\bibitem{susy4}
  T.~Roy and M.~Schmaltz,
  arXiv:hep-ph/0509357.

\bibitem{susy5}
  C.~Csaki, G.~Marandella, Y.~Shirman and A.~Strumia,
  arXiv:hep-ph/0510294.

\bibitem{flavon}
  F.~Bazzocchi, S.~Bertolini, M.~Fabbrichesi and M.~Piai,
  Phys.\ Rev.\ D {\bf 68}, 096007 (2003)
  [arXiv:hep-ph/0306184].
  
\bibitem{quint}
  C.~T.~Hill and A.~K.~Leibovich,
  Phys.\ Rev.\ D {\bf 66}, 075010 (2002)
  [arXiv:hep-ph/0205237].

\bibitem{inflate}
  N.~Arkani-Hamed, H.~C.~Cheng, P.~Creminelli and L.~Randall,
  JCAP {\bf 0307}, 003 (2003)
  [arXiv:hep-th/0302034].
   
}
\end{thebibliography}
\end{document}